\documentclass[twocolumn]{aastex63}
\usepackage{blindtext}
\usepackage{amsmath,amssymb}
\usepackage{gensymb}
\usepackage{appendix}
\usepackage{physics}
\usepackage{multirow}
\usepackage{textcomp}
\usepackage{array}
\usepackage{threeparttable}
\usepackage{CJK}
\usepackage{booktabs}

\received{}
\revised{}
\accepted{}
\shorttitle{Heat-Flux Limited Moist Convection on Ice Giants}
\shortauthors{}
\graphicspath{{./}{figures/}}

\begin{document}
\begin{CJK}{UTF8}{gbsn}

\title{Heat-Flux Limited Cloud Activity and Vertical Mixing in Giant Planet Atmospheres with an Application to Uranus and Neptune}

\correspondingauthor{Huazhi Ge}
\email{huazhige@caltech.edu}

\author[0000-0001-6719-0759]{Huazhi Ge (葛华志)} 
\affil{51 Pegasi b Fellow}
\affil{Department of Earth and Planetary Sciences, University of California, Santa Cruz, 1156 High Street, Santa Cruz, CA 95064, USA}
\affil{Division of Geological and Planetary Sciences, California Institute of Technology, 1200 E California Blvd, Pasadena, CA 91125, USA}

\author[0000-0002-8280-3119]{Cheng Li}
\affiliation{Department of Climate and Space Sciences and Engineering, University of Michigan, Ann, Arbor, M 48109, USA}

\author[0000-0002-8706-6963]{Xi Zhang}
\affiliation{Department of Earth and Planetary Sciences, University of California, Santa Cruz, 1156 High Street, Santa Cruz, CA 95064, USA}

\author[0000-0002-6293-1797]{Chris Moeckel}
\affiliation{Department of Earth and Planetary Science, University of California, Berkeley, 307 McCone Hall, Berkeley, CA 94720, USA}

\begin{abstract}

Storms operated by moist convection and the condensation of $\rm CH_{4}$ or $\rm H_{2}S$ have been observed on Uranus and Neptune. However, the mechanism of cloud formation, thermal structure, and mixing efficiency of ice giant weather layers remains unclear. In this paper, we show that moist convection is limited by heat transport on giant planets, especially on ice giants where planetary heat flux is weak. Latent heat associated with condensation and evaporation can efficiently bring heat across the weather layer through precipitations. This effect was usually neglected in previous studies without a complete hydrological cycle. We first derive analytical theories and show the upper limit of cloud density is determined by the planetary heat flux and microphysics of clouds but independent of the atmospheric composition. The eddy diffusivity of moisture depends on the heat fluxes, atmospheric composition, and gravity of the planet but is not directly related to cloud microphysics. We then conduct convection- and cloud-resolving simulations with SNAP to validate our analytical theory. The simulated cloud density and eddy diffusivity are smaller than the results acquired from the equilibrium cloud condensation model and mixing length theory by several orders of magnitude but consistent with our analytical solutions. Meanwhile, the mass-loading effect of $\rm CH_{4} $ and $\rm H_{2}S$ leads to superadiabatic and stable weather layers. Our simulations produced three cloud layers that are qualitatively similar to recent observations. This study has important implications for cloud formation and eddy mixing in giant planet atmospheres in general and observations for future space missions and ground-based telescopes.

\end{abstract}

\section{Introduction} \label{sec:introduction}

The ice giants, Uranus and Neptune, are among the least understood regimes in the solar system due to their distance from the Earth. Uranus is particularly noteworthy for its near-zero planetary heat flux \citep{pollack1986estimates} and its relatively tranquil and hazy atmosphere. In contrast, Neptune, which is farther from the sun and receives less insolation than Uranus, has a higher observed planetary heat flux \citep{pollack1986estimates} and a more dynamic atmosphere characterized by intermittent storms and dark spots compared to Uranus. 

Since the flyby of Voyager 2 in the 1980s, observers have provided plentiful details about cloud activity in ice giants. Cloud activities and intermittent storms, which are direct evidence of active moist convection in ice giant atmospheres, have been observed on Uranus \citep[e.g.,][]{sromovsky2000ground,de2015record,irwin2016spectral} and Neptune \citep[e.g.,][]{hammel1989discrete,sromovsky1993dynamics,gibbard2003altitude,molter2019analysis,chavez2023evolution}. Radio occultations from Voyager 2 suggest that $\rm CH_{4}$ condenses at the 1-to-2 bar region and, therefore, the major composition of clouds in this region should be $\rm CH_{4}$ \citep{lindal1987atmosphere,lindal1992atmosphere}. Observations suggest that the S/N ratio of ice giant atmosphere is significantly larger than one and the solar value by directly detecting $\rm H_{2}S$ but not $\rm NH_{3}$ in their atmospheres \citep{irwin2018detection,irwin2019probable}. Therefore, the deep cloud deck at about 5 bar is more likely composed of $\rm H_{2}S$ cloud instead of $\rm NH_{3}$ \citep{irwin2022hazy}. Observations also suggest that there are high-altitude $\rm CH_{4}$ clouds near the tropopause at 0.1 bar on ice giants, but they are more commonly seen and active on Neptune than Uranus \citep[e.g.,][]{gibbard2003altitude,irwin2022hazy,chavez2023evolution}. Summaries of cloud activities and storm events can be found in recent review papers \citep[e.g.,][]{hueso2019atmospheric,hueso2020convective,fletcher2022atmosphere,palotai2022moist}.

Clouds have a significant contribution to the opacity and albedo of ice giant atmospheres. Visible to near-infrared observations can help constrain the cloud structure and evolution from cloud opacity and spectroscopy on Uranus and Neptune \citep[e.g.,][]{hueso2019atmospheric,irwin2022hazy,chavez2023evolution} and, therefore, provide useful information to help theorists understand and constrain the atmospheric dynamics on ice giants. 

From the theoretical aspect, cloud activity and moist convection are expected to be unique on ice giants since the exceedingly abundant cloud-forming species \citep[i.e., possibly around 30-to-80 times solar for $\rm CH_{4}$ and $\rm H_{2}S$ according to][]{baines1995abundances,sromovsky2011methane,karkoschka2011haze,molter2021tropospheric} are heavier than the hydrogen-helium background. Linear perturbation theory predicts that the condensation of $\rm CH_{4}$ can inhibit convection and alter the temperature profiles to superadiabatic due to the mass loading effect \citep{guillot1995condensation,leconte2017condensation,friedson2017inhibition}. Radiation may take over the heat transport at the superadiabatic layer from convection and reduce convective mixing \citep{leconte2017condensation}. On the other hand, \citet{ackerman2001precipitating} proposed that latent heating and cooling induced by condensation and evaporation could play a vital role in heat transport, leading to reduced convective heat flux and decreased mixing efficiency in the weather layer. However, this effect was not eventually included in their cloud model and the other cloud models \citep{ohno2017condensation}. While the relative importance between the latent heat effect and mass loading effect remains unclear, it is anticipated that the mixing efficiency of moisture in the weather layer of ice giants would be reduced by one or both of these effects. Consequently, fully convective atmosphere models tend to overestimate mixing and subsequent cloud abundance by neglecting these effects.

Moist convection in the observable weather layers of ice giants, along with the coupled cloud activities, moisture mixing efficiency or eddy diffusivity, and temperature structure, has not yet been systematically investigated. Current one-dimensional (1D) giant planet cloud models either focus on the fresh cloud content inside localized strong updrafts or storms by neglecting precipitations \citep[e.g.,][]{weidenschilling1973atmospheric,wong2015fresh,atreya2020deep,hueso2020convective} or they do not include the non-local latent heat transport induced by phase transition \citep[e.g.,][]{ackerman2001precipitating,ohno2017condensation}. Thus, there is a need for more comprehensive studies to improve our understanding of the critical aspects of weather in ice giant atmospheres before future ice giant missions.

In this study, we aim to estimate the upper bounds of the globally averaged cloud density and effective eddy diffusivity of moisture in giant planet atmospheres by taking a complete hydrological cycle of moisture with both condensation of clouds and evaporation of precipitation and their contribution to heat transport into account. In particular, this study focuses on the cloud formation and eddy diffusion of $\rm CH_{4}$ and $\rm H_{2}S$ in Uranus and Neptune. First, we use the order of magnitude estimation starting from first-principle continuity equations (i.e., conservation of mass and energy) to show that moist convection is limited by the latent heat flux and planetary heat flux of ice giants. We present analytical estimations of the cloud density and eddy diffusivity in Section~\ref{sec:cloud} and \ref{sec:kzz}, respectively. Then, in Section~\ref{sec:CRM-model}, we employ the nonhydrostatic model, SNAP, which can consistently resolve turbulence and thermal convection, to study the temperature structure, cloud density, and mixing efficiency. We show that the numerical solutions from the cloud-resolving and convection-resolving simulations are consistent with the analytical estimations. Finally, we provide discussions about the difference between Uranus and Neptune and implications for the other giant planets, future observation, and ice giant missions in Section~\ref{sec:discussions}. We conclude take-away points in Section~\ref{sec:conclusion}.

\section{Constrain Cloud Content and Mixing Efficiency in Giant Planet Weather Layers}
\label{sec:self-regulation}

\begin{figure*}[t]
\centering
\includegraphics[width=\linewidth]{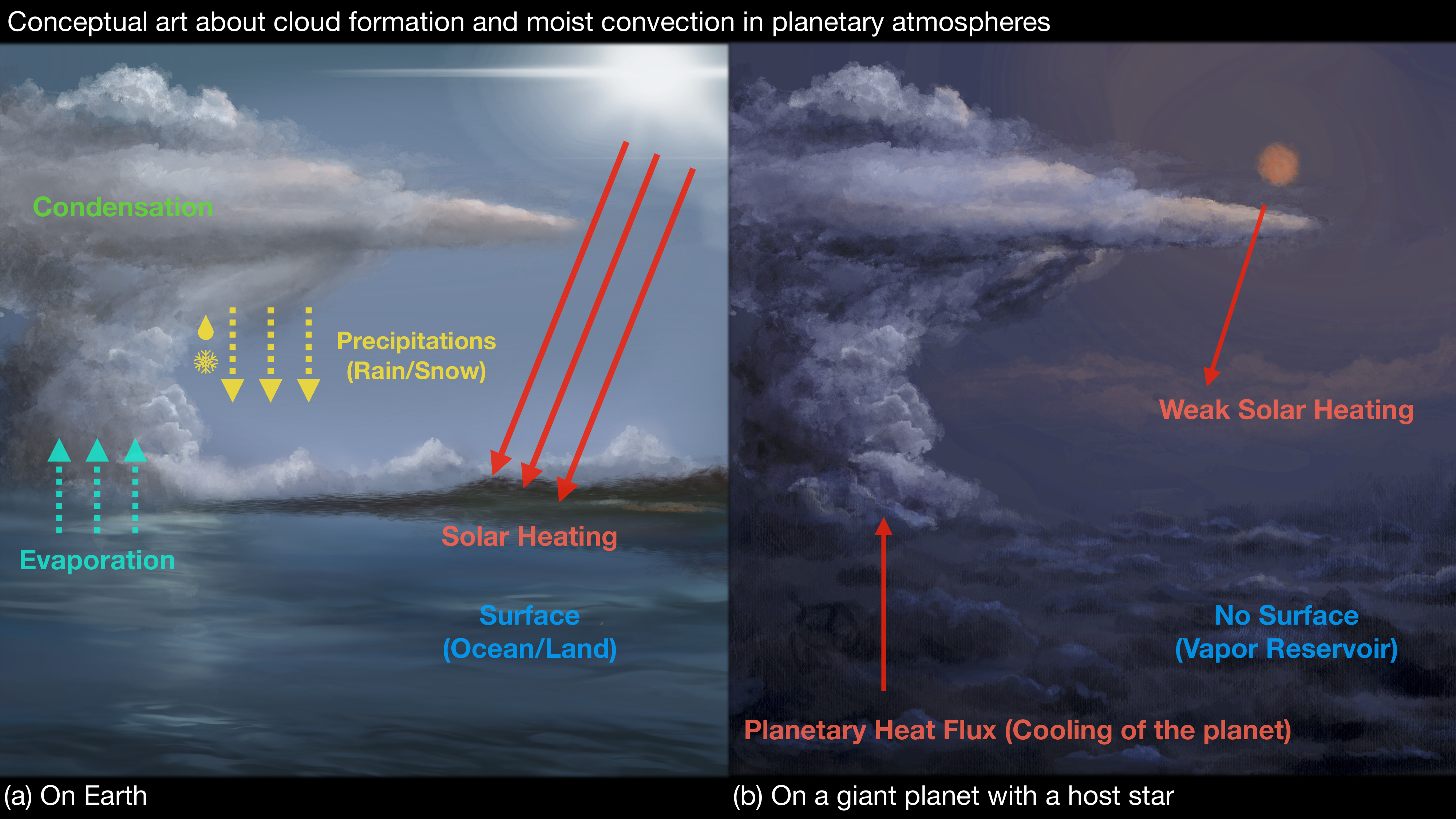}
\caption{Conceptual art about cloud formation and moist convection in the atmosphere of different planets. Panel (a) shows that solar radiation is the driving force and the energy source of cloud formation in Earth's atmosphere. On the other hand, panel (b) shows that the planetary heat flux is the driving force of cloud formation on giant planets since solar radiation stabilizes the atmosphere and against convection. Note that the concept in (b) is also applicable to free-floating substellar bodies like field brown dwarfs but not applicable to strongly irradiated giant planets like hot or warm Jupiter. Cloud morphology in this art does not necessarily represent the real ones.}
\label{Fig-concept}
\end{figure*}

On Earth, solar radiation drives moist convection and cloud formation by warming the surface (Fig~\ref{Fig-concept}a). However, the situation is fundamentally different on giant planets. Solar radiation warms and then stabilizes the upper atmosphere instead of driving convection on giant planets. The subsequent stabilization hinders turbulent mixing driven by buoyancy at the top of the convective zone. As a result, cloud activity on giant planets must be controlled by distinct mechanisms compared to the solar radiation in the Earth's atmosphere. Specifically, gas giant cloud formation is primarily controlled by the planetary heat flux \citep{ackerman2001precipitating} and the cooling of the planet, which, in turn, is related to the temperature at the radiative zone near the top of the atmosphere (Fig~\ref{Fig-concept}b).

This section aims to derive analytical solutions from the first principles to estimate the upper limits of averaged cloud density and mixing efficiency that can be generalized for all giant planets. Then, we test these theories on Uranus and Neptune, where planetary heat fluxes are comparatively small compared to the other observable giant planets in the universe \citep{pollack1986estimates} and whose upper cloud structures and properties of aerosol particles have been recently constrained by observations \citep{irwin2022hazy}. 

To achieve this, we first define some terms. The averaged quantities, either cloud density or moisture mixing efficiency, are referred to as globally averaged numbers smoothed over a timescale that is considerably longer than meteorological or climatic variations \citep[e.g., decades,][]{chavez2023evolution} and yet shorter than geological timescales (i.e., millions of years). To be clear, we are not investigating the cloud structure in any transient and localized storms. By doing so, we can neglect complications arising from localized intermittent storms, energy loss due to the Kelvin-Helmholtz contraction of the planet, and mass loss of the atmosphere due to the atmospheric escape. 

Moist convection, which produces clouds, must be subjected to the conservation of energy and mass and follow continuity equations. Specifically, the conservation of energy means the averaged net heat flux, resulting from various heat transport mechanisms such as phase transition, turbulent eddy convection, thermal diffusion, and radiation, remains almost constant in the vertical direction of each atmospheric layer by neglecting the geometric expansion of the atmosphere in the radial direction. This net heat flux is equivalent to the observed planetary heat flux at the top of the atmosphere with solar radiative heating excluded. If the net heat flux of a certain layer is larger or smaller than the planetary heat flux over a long period, it will lead to continuous warming and cooling at that layer which can be considered the climatic variation and therefore violates the assumption we made before. On the other hand, the mass conservation of total clouds can be interpreted as the production rate of clouds being equivalent to the loss rate of clouds at a statistically steady state, which directly relates clouds to precipitations.

To simplify the problem, we first neglect some trivial terms that have limited contributions to heat transport (e.g., kinetic energy and thermal diffusion). The averaged net heat flux generated by phase transition, convection, and radiation is equivalent to the observed planetary heat flux at the top of the atmosphere,

\begin{equation}
    F_{\rm latent} + F_{\rm conv} + F_{\rm rad} \sim F_{\rm rad}^{\rm TOA}.
\end{equation}
where $F_{\rm latent}$ is associated with the phase transition of cloud-forming species and condensates, $F_{\rm conv}$ is the convective heat flux, $F_{\rm rad}$ is the radiative flux at each layer, and $F_{\rm rad}^{\rm TOA}$ is the emission flux observed at the top of the atmosphere (TOA) but excludes the incoming solar radiation and is usually interpreted as the internal or planetary heat flux. Latent heat flux, $F_{\rm latent}$, requires settling precipitation to bring condensates downward into the deep warm atmosphere to complete a complete hydrological cycle and recycle the latent heat released during the condensation. Therefore, equilibrium cloud condensation models (ECCM) \citep[e.g.,][]{weidenschilling1973atmospheric} and large-scale general circulation models \citep[e.g.,][]{lian2010generation} without precipitation cannot capture the physics of latent heat flux. 

Since the averaged convective heat flux, $F_{\rm conv}$, and radiative heat flux, $F_{\rm rad}$, are usually positive below the photosphere, we can immediately conclude that $F_{\rm latent} < F_{\rm rad}^{\rm TOA}$. 

In the weather layer, the mass flux of downward precipitation is equivalent to the net mass flux of moisture, which is the mass conservation of all cloud-forming species. Then, the mass conservation of these species can be written in a flux form (see derivations in Appendix~\ref{app:mass-flux}),

\begin{equation}
    \overline{\rho_{v}w} \sim -\overline{\rho_{p}V_{T}},
\end{equation}
where $\overline{A}$ is the global and long-term average of $A$; $\overline{\rho_{v}w}$ is the net mass flux of the cloud-forming species; $-\overline{\rho_{p}V_{T}}$ is the mass flux of precipitations; $\rho_v$ is the vapor density; $w$ is the vertical velocity of the gas; $\rho_p$ is the density of precipitation; and $V_{T}$ is the terminal velocity, known as the sedimentation velocity, of precipitations. Here, we assume the net flux of precipitations due to the background convection, $\overline{\rho_{p} w}$, is negligible.

The latent heat flux, $F_{\rm latent}$, is related to the mass flux of precipitations, $\overline{\rho_{p}V_{T}}$, by $F_{\rm latent} \sim \overline{\rho_{p}V_{T}}L_{v}$, where $L_{v}$ is the specific latent heat. An explanation of this equation is that the upward vapor flux possesses the larger chemical potential (i.e., latent heat), and the downward condensate flux carries the smaller chemical potential. The net exchange of two streams results in the latent heat flux. One can also derive this relation from the first law of thermodynamics \citep{li2019simulating}.

Now, we can estimate the upper limit of the averaged cloud density and mixing efficiency.

\subsection{Upper Limit of Cloud Column Density and Cloud Density} 
\label{sec:cloud}

The upper limit of cloud density can be estimated from the mass balance of precipitations. Cloud particles aggregate into large precipitable condensates, such as raindrops, through collision and coalescence until they are large and heavy enough to fall. All precipitable condensates are originally formed by the condensation of vapor and eventually evaporate in giant planet atmospheres. On a long-term average, the production rate of precipitations is equivalent to the mass flux of precipitation that leaves the cloud base. 

The production rate of precipitation of a vertically integrated cloud column in $\rm kg\;m^{-2}\;s^{-1}$ is,

\begin{equation}
    \int_{\rm cloud\;bot}^{\rm cloud\;top} \frac{\rho_{c}}{\tau_{\rm c \rightarrow p}} dz 
    \sim \frac{\sigma_{c}}{\tau_{\rm c \rightarrow p}}
    \sim \frac{\rho_{c}H_{c}}{\tau_{\rm c \rightarrow p}}
\end{equation}
where $\rho_{c}$ is cloud density in $\rm kg\; m^{-3}$; $\sigma_{c}$ is the vertically integrated cloud column density, $\int_{\rm cloud\;bot}^{\rm cloud\;top} \rho_{c} dz$, in $\rm kg\; m^{-2}$; $\tau_{\rm c\rightarrow p}$ is the cloud lifetime that measures how fast the cloud particles can grow or aggregate into precipitable raindrops/snowflakes/hailstone, hereafter defined as the autoconversion timescale, which is prescribed as a constant in this study; $H_{c}$ is the thickness of the cloud layer. The first approximation assumes the autoconversion timescale is constant inside the cloud column, and the second approximation further assumes the cloud is uniformly distributed inside the column in which the cloud density is constant in altitude.

Then, we can estimate the cloud column density and cloud density from the planetary heat flux. The column production rate of precipitations in $\rm kg\; m^{-2}\; s^{-1}$ is equivalent to the mass flux of precipitations (i.e., the column loss rate of clouds in $\rm kg\; m^{-2}\; s^{-1}$) because of the mass conservation of precipitation at the statistically steady state,

\begin{equation}
    \frac{\sigma_{c}}{\tau_{\rm c \rightarrow p}} \sim \frac{\rho_{c}H_{c}}{\tau_{\rm c \rightarrow p}} \sim -\overline{\rho_{p}V_{T}} .
\end{equation}

On the other hand, the mass flux of precipitation is determined by the latent heat flux, which must be smaller than the planetary heat flux in the long-period average,

\begin{equation}
    \frac{\sigma_{c}}{\tau_{\rm c \rightarrow p}} \sim \frac{\rho_{c}H_{c}}{\tau_{\rm c \rightarrow p}} \sim -\overline{\rho_{p}V_{T}} \sim \frac{F_{\rm latent}}{L_{v}} < \frac{F_{\rm rad}^{\rm TOA}}{L_{v}}.
\end{equation}
Hence, we can constrain the cloud column density by the planetary heat flux and autoconversion timescale,

\begin{equation}
\label{eq:cloud-column-limit}
\begin{split}
    \sigma_{c} < \frac{F_{\rm rad}^{\rm TOA}\tau_{\rm c \rightarrow p}}{L_{v}} 
\end{split}
\end{equation}

Now, we can plug in some numbers and estimate the cloud column density of $\rm CH_{4}$ and $\rm H_{2}S$ clouds on ice giants. The internal heat flux of Uranus and Neptune is less than 0.5 $\rm W\;m^{-2}$ \citep{pollack1986estimates,scheibe2019thermal}. The latent heat of $\rm CH_{4}$ and $\rm H_{2}S$ is about $\rm \sim 5\times 10^{5} \; J \; kg^{-1}$ (i.e., enthalpy of vaporization, one can find experiments constrained numbers on \href{https://webbook.nist.gov/chemistry/form-ser/}{NIST Chemistry WebBook}). The autoconversion timescale can be constrained by observations and previous microphysics works on ice giants. Cloud microphysics studies suggest the faster timescale among coalescence and coagulation processes is about $\sim 10^2$-to-$10^4$ s for cloud particles with the size ranging from sub-microns to sub-meters \citep{carlson1987cloud,palotai2022moist}. Indeed, these timescales are sensitive to the collision rate of particles inside the cloud and, therefore, sensitive to the cloud density itself \citep[Eq.3-8 in][]{carlson1987cloud}. With a reduced cloud density, the coagulation-coalescence timescale suggested by \cite{carlson1987cloud} and \cite{palotai2022moist} should be longer. From the observational side, high albedo clouds can last for about ten days \citep{sromovsky2005dynamics,de2015record,molter2019analysis}. Here, we use $10^4$ s (i.e., a few hours) for the autoconversion timescale to estimate the cloud column density and cloud density, but the real timescale might change based on $\sim 10^4$ s by one or two orders of magnitude.

We can then obtain the upper limit of the cloud column density by assuming the $\tau_{\rm c \rightarrow p}\sim 10^{4}$ s, $L_{v}\sim 5\times 10^5\;{\rm J\;kg^{-1}}$, and $F_{\rm rad}^{\rm TOA}\sim 0.5 \; {\rm W\;m^{-2}}$. The cloud column density, in this case, is about,

\begin{equation}
    \sigma_{c} < \frac{0.5 {\rm \;W\;m^{-2}\;} \times 10^{4}\;{\rm s}}{5\times 10^{5} {\rm \;J\;kg^{-1}}} \sim 10^{-2} \rm{\;kg\;m^{-2}}.
\end{equation}

Previous studies usually present their results in cloud density or cloud content in $\rm kg\;m^{-3}$ \citep[e.g.,][]{weidenschilling1973atmospheric,hueso2020convective}. To compare with previous results, we estimate the upper limit of the cloud density by assuming the cloud density is constant in the vertical direction,

\begin{equation}
\label{eq:cloud-limit}
    \rho_{c} < \frac{F_{\rm rad}^{\rm TOA}\tau_{\rm c \rightarrow p}}{L_{v}H_{c}}.
\end{equation}
Here, we adopt the cloud thickness from previous cloud models and numerical solutions presented in later sections (Fig~\ref{Fig5-cloud}) to estimate the typical cloud density. Previous 1D cloud models \citep[e.g.,][]{weidenschilling1973atmospheric,hueso2020convective} suggest the $\rm CH_{4}$ cloud density changes by a factor of a hundred from about $\rm \sim$ 2 bars to 0.5 bars for 80 times solar on Uranus and Neptune. The cloud-resolving simulations suggest the cloud thickness is also about one to a few scale heights with varying cloud density in the vertical (Fig~\ref{Fig5-cloud}). Here, in order to make the cloud density comparable to previous works, we assume the cloud thickness is slightly smaller than the pressure scale height. We set the cloud thickness to $10^{4}$ m given the local pressure and temperature is about 1 bar and 80 K at the $\rm CH_{4}$ condensation level. 

Thus, the upper limit of the averaged $\rm CH_{4}$ cloud density in ice giant atmospheres is about,

\begin{equation}
    \rho_{c} < \frac{0.5 {\rm \;W\;m^{-2}\;} \times 10^{4}\;{\rm s}}{5\times 10^{5} {\rm \;J\;kg^{-1}}\times 10^{4}\;{\rm m}} \sim 10^{-6} \rm{\;kg\;m^{-3}}.
\end{equation}
For the $\rm H_{2}S$ cloud, this upper limit is similar to the one of $\rm CH_{4}$ since the latent heat of $\rm H_{2}S$ is close to the $\rm CH_{4}$. The autoconversion timescale might be different, but there is a lack of experimental constraints yet to tell us whether this difference is significant. 

Previous cloud models produce a much higher cloud density which is on the level of 0.1-1 $\rm kg\;m^{-3}$ with a vertical extent larger than one scale height \citep[e.g.,][]{weidenschilling1973atmospheric,hueso2020convective}, which is several orders of magnitude larger than the number suggested by Eq~\ref{eq:cloud-limit}.

One can amplify the estimated cloud density by choosing a longer autoconversion timescale with the reduced average cloud density or adopting a smaller cloud thickness. Then, the estimated cloud density from Eq~\ref{eq:cloud-limit} could be amplified by two to three orders of magnitude, with the cloud lifetime being about a month to a year. However, it is still hard to reconcile the six orders of magnitude difference between our estimation and the previous 1D models \citep{weidenschilling1973atmospheric,hueso2020convective}, as it would require a cloud lifetime of more than a century or extremely compact cloud layer within 0.1 m. To address this disparity, we suggest that the cloud density proposed by previous models may be more representative of localized storms or fresh clouds without significant precipitations \citep{wong2015fresh}, where cloud content is supply-limited instead of heat-flux-limited as this study suggested. Extrapolating these numbers to represent the globally averaged cloud density may significantly overestimate the cloud content in ice giant atmospheres. This overestimation might have occurred for all solar system giant planets.

\subsection{The Upper Limit of Mixing Efficiency/Eddy Diffusivity, $K_{zz}$, of Moisture}
\label{sec:kzz}

Now, we estimate the upper limit of the vertical mixing efficiency of moisture. Due to the mass balance of total cloud-forming species, the net mass flux of moisture is equivalent to the mass flux of precipitations, which is also the column loss rate of condensates. Then, the eddy diffusivity of the moisture in $\rm m^{2}\;s^{-1}$ can be directly estimated from the mass flux of precipitations and concentration gradient of moisture \citep[e.g., Chapter 13 in][]{vallis2017atmospheric,zhang2018global},

\begin{equation}
\label{eq:Kzz-general}
\begin{split}
    K_{zz} & \sim \overline{q_{v}w}\Big( \overline{\dfrac{\partial q_{v}}{\partial z}}\Big)^{-1}
           \sim \dfrac{\overline{\rho_{v}w}}{\rho}\Big(\overline{\dfrac{\partial q_{v}}{\partial z}}\Big)^{-1} \\
           & \sim -\dfrac{\overline{\rho_{p}V_{T}}}{\rho}\Big(\overline{\dfrac{\partial q_{v}}{\partial z}}\Big)^{-1} 
           < -\frac{F_{\rm rad}^{\rm TOA}}{\rho L_{v}} \Big( \overline{\frac{\partial q_{v}}{\partial z}} \Big)^{-1}, \\
\end{split}        
\end{equation}
where $q_{v}$ is the mass mixing ratio, or specific humidity, of the moisture. The magnitude of the averaged concentration gradient, $\vert\overline{\partial q_{v}/\partial z}\vert$, determines how much moisture can condense (i.e., be supersaturated) in upwelling motions. A small concentration gradient requires stronger convection with a larger $K_{zz}$ to achieve the same amount of supersaturated vapor compared to an atmosphere with a larger concentration gradient. In an extreme case as a thought experiment, if the concentration is constant in altitude (i.e., zero concentration gradient), no lifted vapor can condense and, therefore, $K_{zz}$ should be infinitely large for any non-zero latent heat flux.

We can then obtain the upper limit of the eddy diffusivity in the cloud region by assuming the background temperature profile is close to dry adiabat and fully saturated,

\begin{equation}
\label{eq:Kzz-estimate}
    K_{zz} < \frac{F_{\rm rad}^{\rm TOA}H_v}{\rho L_{v}\eta_s}, 
\end{equation}
where $\eta_s$ is the volume mixing ratio of the moisture, and  $H_v$ is the scale height and the e-folding length scale of condensing species:

\begin{equation}
    H_{v} \sim \frac{H}{\epsilon} \Big( \frac{\gamma-1}{\gamma}\beta-1 \Big)^{-1}.
\end{equation}
$\epsilon$ is the molecular weight ratio of the moisture to the hydrogen-helium mixture; $H$ is the pressure scale height; $\gamma$ is the polytropic index; and $\beta$ is a dimensionless ratio of latent heat, $L_{v}$, to $R_{v}T$ (see detailed derivations in Appendix~\ref{app-kzz}). The first term in the bracket is related to the latent heat from phase transitions, and the second term is related to the pressure change due to the condensation. 

The scale height of the condensing species is determined by the molecular mass of the moisture and dimensionless latent heat, $\beta$. The typical molecular weight ratio, $\epsilon$, is about 7 (e.g., $\rm H_{2}O$, $\rm NH_{3}$, and $\rm CH_{4}$. $\rm H_{2}S$ is larger), and the number of $\beta$ ranges from 10 to 20. Therefore, $H_{v}$ is generally 15 to 30 times shorter than the pressure scale height, $H$.

Here, we can conclude that the globally averaged vertical mixing efficiency, $K_{zz}$, is primarily determined by three parameters in the weather layer of giant planets. They are the luminosity of the planet (e.g., planetary heat flux) shown as $F_{\rm rad}^{\rm TOA}$, metallicity (e.g., condensing species abundance) shown as $\eta_{s}$, and mass of the planet (e.g., gravity) hiding in the scale height $H$.

Although Eq~\ref{eq:Kzz-estimate} itself is irrelevant to the parameters of cloud microphysics, the evaporation rate and falling speed of condensates might alter the vapor concentration gradient in Eq~\ref{eq:Kzz-estimate} by changing the relative humidity in altitude. 

Plugging some typical numbers of $\rm CH_{4}$ vapor at 0.6 bars into the equation, we can get the upper limit of the eddy diffusivity at the $\rm CH_{4}$ cloud layer of ice giants,

\begin{equation}
\label{eq:slow-rotation}
\begin{split}
    K_{zz,{\rm \;CH_{4}}} & < \frac{0.5}{0.3\times 5\times 10^{5}}\Big[  \frac{10^{-2}}{3\times 10^{4}} \Big( \frac{2\times 12}{7} - 1 \Big) \Big]^{-1} \\
    & \sim 2 \; {\rm m^{2}\;s^{-1}},
\end{split}
\end{equation}
and the upper limit of $K_{zz}$ of $\rm H_{2}S$ vapor at about 3 bars is,

\begin{equation}
\label{eq:fast-rotating}
\begin{split}
    K_{zz,{\rm \;H_{2}S}} & < \frac{0.5}{1\times 5\times 10^{5}}\Big[  \frac{2\times 10^{-3}}{10^{5}} \Big( \frac{2\times 12}{7} - 1 \Big) \Big]^{-1} \\
    & \sim 30 \; {\rm m^{2}\;s^{-1}},
\end{split}
\end{equation}
We can compare these numbers with the eddy diffusivity estimated by the mixing length theory by assuming the convective heat flux dominates the heat transport \citep[e.g.,][]{showman2013atmospheric,wang2015new,zhang2020atmospheric,moses2020atmospheric}. If we focus on the $\rm CH_{4}$ weather layer in ice giants, previous mixing length theories assume the planetary heat flux is completely delivered by buoyancy-driven convection across the weather layer. If we use the pressure scale height to approximate the mixing length, $l \sim H$, and use the thermal expansivity of ideal gases, $1/T$, we get

\begin{equation}
    K_{zz} \sim \Big( \frac{gF_{\rm rad}^{\rm TOA} l^4}{\rho c_{p} T} \Big)^{1/3} \sim 3 \times 10^4 \; {\rm m^{2}\;s^{-1}}.
\end{equation}
Clearly, applying the mixing length theory in the weather layer could greatly overestimate the $K_{zz}$ by neglecting the latent heat flux, which might be the dominant heat transport mechanism. This number might be reduced by a factor of 10-to-$10^2$ by using a shorter mixing length instead of the pressure scale height, but it is still hard to reconcile the result acquired from Eq~\ref{eq:Kzz-estimate}. 

For the fast-rotating case, the scaled $K_{zz}$ is about (i.e., in high latitudes),

\begin{equation}
\begin{split}
    K_{zz} & \sim \frac{gF_{\rm rad}^{\rm TOA}}{\rho c_{p} T \Omega^2} \sim \frac{10\times 0.5}{0.3\times 10^4 \times 10^2 \times (10^{-4})^2} \\ &
    \sim 10^3 \; {\rm m^{2}\;s^{-1}},
\end{split}
\end{equation}
where the length scale is approximated by $l\sim w/\Omega$, with $\Omega$ representing the rotation rate of the planet \citep[see more detailed discussions in Section 2.2 of][]{wang2015new}. However, the mixing efficiency derived from this approximation is still three orders of magnitude larger than the results obtained from Eq~\ref{eq:Kzz-estimate}. 

Therefore, it is evident that the dynamical feedback of latent heat flux, controlled by the full hydrological cycles of cloud-forming species with sedimentations, plays a crucial role in heat transport within ice giant weather layers. Consequently, a careful treatment of the latent heat flux is essential when estimating the mixing efficiency or eddy diffusivity of moisture, $K_{zz}$, in giant planet weather layers.

In which situation can we apply this theory? From the comparison to mixing length theories, we can see that Eq~\ref{eq:Kzz-estimate} is useful to weather layers where phase transition dominates heat transport in weakly forced atmospheres. $\rm CH_{4}$ and $\rm H_{2}S$ weather layers in ice giants clearly satisfy this criterion. It is important to note that this estimation only applies to the main cloud layer but not to regions with very low saturation vapor pressure (i.e., low abundance) since free convection dominates heat transport there. If one applies Eq~\ref{eq:Kzz-estimate} to estimate the eddy diffusivity in the region controlled by free convection, it would then provide a very large mixing efficiency which is significantly larger than the number estimated by the mixing length theory (i.e., Eq~\ref{eq:slow-rotation} or Eq~\ref{eq:fast-rotating}) and thus is not useful as an upper bound. For the mixing efficiency of $\rm H_{2}S$ on Neptune, this threshold occurs at about $\sim 2$ bars.

It is also noticeable that this estimation has nothing to do with the mass-loading effect induced by the molecular weight difference between moisture and dry components. The mixing of the atmosphere is determined by the dominant heat transport mechanism (e.g., radiation versus convection), but the compositional gradient is usually not directly related to the heat transport. According to the Schwarzschild-Ledoux criterion, an atmosphere with a molecular weight gradient could be convective with a superadiabat. In giant planet atmospheres, the compositional gradient controlled by the mass-loading effect could significantly reduce the mixing efficiency if it opens a superadiabatic radiative zone as suggested by \cite{guillot1995condensation}, \cite{leconte2017condensation}, \cite{friedson2017inhibition}, and \cite{markham2022convective}. 

One may also adopt this estimation to the quenching of any chemical with significant chemical potential change, for example, the transition of para- and ortho-hydrogen and formation of $\rm NH_{4}HS$ cloud if the heat flow is also primarily carried out by the chemical potential instead of free convection.

\section{Comparison to Convection-Resolving and Cloud-Resolving Simulations}
\label{sec:CRM-model}

In this section, we show the results of the nonhydrostatic cloud-resolving simulations. We use SNAP \citep{li2019simulating,ge2020global} to simulate the cloud structure, mixing efficiency, and temperature structure in ice giant atmospheres. SNAP is built on top of $\rm Athena^{++}$, a magnetohydrodynamic code employing the finite volume method (FVM) for astrophysics \citep{stone2020athena++}. SNAP incorporates several numerical schemes to boost the efficiency and accuracy of simulating turbulent convection in the regime of atmospheric science. Most important, SNAP directly solves nonhydrostatic density and pressure, and, therefore, the buoyancy of eddies, as prognostic variables to directly resolve convection with the presence of compositional and thermal variations.

We utilize the $\rm 5^{th}$-order accurate weighted essentially non-oscillatory scheme (WENO5) \citep{shu1998essentially} for reconstruction inside the discretized cells and low Mach number Riemman solver (LMARS) \citep{chen2018towards} to solve the Riemann problem on the faces between cells. We also enable the vertical-implicit-correction scheme to improve computational efficiency \citep{ge2020global}. We incorporate hydrological systems of $\rm CH_{4}$ and $\rm H_{2}S$ with simplified microphysics \citep{li2019simulating} to study the thermal structure, cloud formation, and mixing efficiency in the weather layer of ice giants. 

We focus on Neptune for numerical simulations because Uranus is difficult to simulate due to its intrinsic properties, for example, comparatively smaller heat flux (i.e., leading to a significantly longer convective timescale) and smaller surface gravity (i.e., leading to a larger domain due to the extended scale height). However, if the numerical solutions of Neptune support and are consistent with the analytical solutions discussed in previous sections, we can then use the analytical method to estimate the cloud activity and $K_{zz}$ of the moistures on Uranus.

\subsection{Methods --- Model Setup}
\label{sec:methods}

(1) \textbf{Domain} We adopt 2D simulations in this study, similar to previous studies on Jupiter \citep{nakajima2000numerical,sugiyama2011intermittent,sugiyama2014numerical}. 3D simulations with high spatial resolutions can resemble more realistic convection in the atmosphere. But, they are computationally expensive because cloud-resolving simulations of giant planets take a long time to converge --- the radiative cooling timescale of ice giants is about century-long \citep{li2018high}. Thus, we adopt 2D simulations in this study for better computational efficiency. We employ similar 2D setups in previous works \citep{sugiyama2011intermittent,sugiyama2014numerical,li2019simulating}. We conduct simulations in a 2D localized box with 350 km in the vertical direction in a height coordinate system and 500 km horizontal with a vertical spacing of 1.75 km and horizontal spacing of 5 km. The pressure ranges from about 140 bars to 0.6 mbars. The lower boundary is deep enough to ensure convection at the weather layer is not affected by boundary effects.

We employ the surface gravity of Neptune, which is 11.15 $\rm m\;s^{-2}$, for all simulations. Uranus has a slightly smaller gravity (i.e., 8.8 $\rm m\;s^{-2}$) than Neptune's, and the analytical estimation in Section~\ref{sec:kzz} shows that only the eddy diffusivity, $K_{zz}$, should be slightly reduced with a lower gravity since the eddy diffusivity is linearly proportional to the pressure scale height.

(2) \textbf{Internal heating and radiative cooling} The observed internal heat flux of Uranus is almost zero, which is hard to incorporate into simulations to study the cloud formation, vertical mixing, and thermal structure because heating and cooling from the heat flux are mandatory for moist convection. On the contrary, Neptune's internal heat flux is around 0.4 $\rm W\;m^{-2}$, which is much larger and technically feasible for simulations \citep{pollack1986estimates,scheibe2019thermal}. Thus, we utilize an internal heat flux of 0.5 $\rm W\;m^{-2}$ in our simulations, which is the upper limit of Neptune's planetary heat flux. We implement a radiative cooling layer above 1 or 2 bars with a universal and constant body cooling rate to balance the energy budget in the domain. The cooling rate represents the net radiative cooling, and we do not specify solar heating in the simulations. To achieve convergence faster, we initialize simulations with a much higher radiative cooling rate and internal heat flux during the first 20,000 days (24 simulation hours). The initial heating and cooling fluxes are set as 5 $\rm W\;m^{-2}$. After the spin-up phase, we run another 15,000 days with heating and cooling fluxes equal to 0.5 $\rm W\;m^{-2}$ to achieve the steady state for analysis. 

(3) \textbf{Composition} Yet, we do not know the accurate composition of Uranus' and Neptune's atmospheres. Spectroscopic observations suggest that the $\rm CH_{4}$ abundance should be around 80 $\pm$ 20 times solar \citep[e.g.,][]{baines1995abundances,sromovsky2011methane,karkoschka2011haze,atreya2020deep}. We simulate the 2D atmosphere with 10 times solar, 30 times solar, and 50 times solar $\rm CH_{4}$ and $\rm H_{2}S$. We use the solar value suggested by \cite{asplund2009chemical}. We initialize the simulation with a chemical equilibrium moist adiabat, in which the concentration of moisture follows the saturation curve \citep{li2018moist}, which is similar to the adiabat derived from the equilibrium cloud condensation model (ECCM) but without cloud \citep{weidenschilling1973atmospheric}. We set the $\rm He/H_{2}$ ratio as 0.131 \citep{irwin2022hazy}. We relax the concentration of moisture to the initial value with a timescale of $10^4$ s at the bottom to mimic the mixing in the deep atmosphere. 

(4) \textbf{Cloud Microphysics} The microphysics of cloud particles and precipitating condensates are controlled by three constant parameters in simulations, autoconversion rate (one over the autoconversion timescale, $\tau_{\rm c \rightarrow p}$), evaporation rate, and settling velocity of precipitations. We exclude the possibility of supersaturation in simulations by assuming the supersaturated moisture condenses in one numerical time step (about 10 seconds) since condensation timescale is generally less than 10 s to form particles smaller than 1 $\rm \mu m$ \citep{carlson1987cloud,palotai2022moist}. We set the autoconversion rate as $\rm 10^{-4}\;s^{-1}$. This number is inferred from previous cloud microphysics models \citep{carlson1987cloud,palotai2022moist} and observations \citep{sromovsky2005dynamics,de2015record}. We have presented the reasons for choosing this number in Section~\ref{sec:cloud}. The sedimentation velocity and evaporation rate determine how far the precipitations can reach. On ice giants, precipitations might be a combination of snowflakes, raindrops, and hailstones. Snowflakes usually precipitate slowly on the order of 1 $\rm m\;s^{-1}$ \citep[Fig 9 in][]{palotai2022moist}, raindrops precipitate at about 10 $\rm m\;s^{-1}$ \citep{carlson1987cloud}, and hailstones precipitate fast with the falling speed possibly exceeding 100 $\rm m\;s^{-1}$ \citep{guillot2020storms}. Hailstones might be formed in strong updrafts, which are identified as storms in observations, while snowflakes may form in tranquil regions. Here, we assume the precipitating $\rm CH_{4}$ and $\rm H_{2}S$ particles have a universal shape and size with the same sedimentation velocity, and it is 10 $\rm m\;s^{-1}$, which is a number between two cases. We do not find a reasonable constraint for the evaporation timescale of $\rm CH_{4}$ and $\rm H_{2}S$ condensates from the current ice giant cloud microphysics models. Therefore, we use a rough e-folding evaporation timescale inferred from the falling distance (i.e., $\tau_{\rm eva}\sim H_{\rm fall}/V_{T}$) of $\rm H_{2}O$ raindrops in Jupiter and Saturn \citep[e.g., Table 2 in][]{loftus2021physics}. Then we convert the inferred timescale to the evaporation rate, $1/\tau_{\rm eva}$, for this study. The corresponding evaporation rate is then about $\rm 3\times 10^{-2}\;s^{-1}$.

\begin{figure*}[t]
\centering
\includegraphics[width=\linewidth]{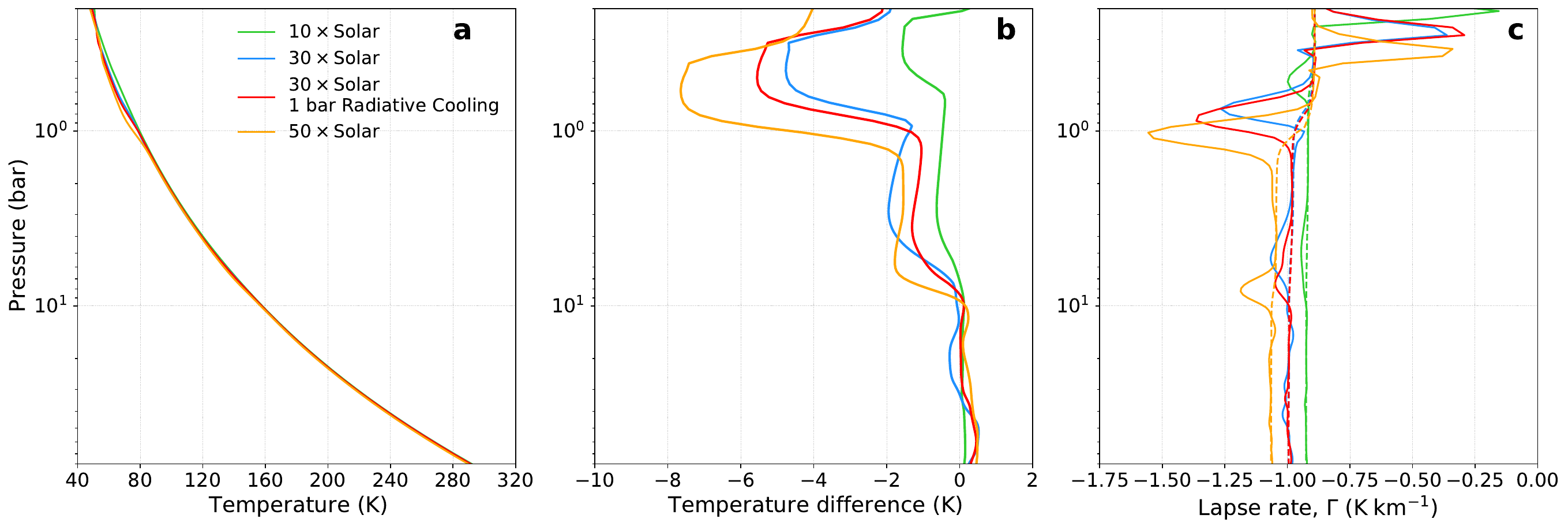}
\caption{Horizontally averaged temperature profile (panel \textbf{a}), temperature difference to the initial chemical equilibrium moist adiabat (panel \textbf{b}), local atmospheric lapse rate (solid lines in panel \textbf{c}), and local adiabatic lapse rate (dashed lines in panel \textbf{c}) with different metallicity. We first collect the physical quantity (i.e., temperature, lapse rate) of every timestep. The local adiabatic lapse rate is dry adiabatic if the relative humidity is smaller than 100\% (i.e., subsaturated) or moist adiabatic if the relative humidity is 100\% (i.e., saturated). Then, we compute the averaged value as a function of pressure or height. Pressure and altitude are interchangeable at the statistically steady state in local simulations since the spatial variation of pressure at a certain altitude is negligible.}
\label{Fig1-T-P}
\end{figure*}

\section{Results of Cloud-Resolving Simulations}\label{sec:result}

\subsection{Superadiabat and Atmospheric Stability at $\rm CH_{4}$ and $\rm H_{2}S$ Weather Layers}
\label{sec:T-N^2}

\begin{figure*}[t]
\centering
\includegraphics[width=\linewidth]{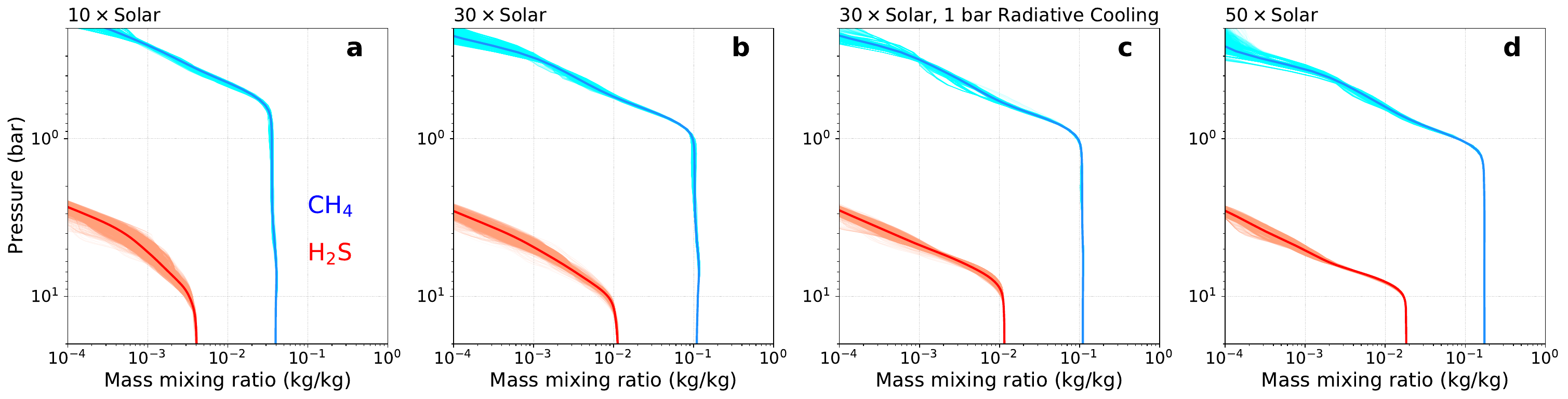}
\caption{Horizontally averaged mass mixing ratio (i.e., specific humidity) of $\rm CH_{4}$ (blue lines) and $\rm H_{2}S$ (red lines) vapor and snapshots with temporal and spatial variations (shaded region). Note that snapshots do not collect the specific humidity at every timestep and may miss the variation caused by short-lived intermittent storms.}
\label{Fig2-abundance}
\end{figure*}

Fig~\ref{Fig1-T-P}\textbf{a} shows horizontally averaged temperature profiles in simulated Neptune's troposphere with different metallicity. We are interested in the simulated temperature profile and how much they deviate from local adiabats. Temperature profiles are widely used for radio observations to retrieve the $\rm H_{2}S$ abundance as a function of pressure and latitude in ice giants. They are usually assumed to be dry or moist adiabatic \citep[e.g.,][]{molter2019analysis,tollefson2019neptune,tollefson2021neptune,molter2021tropospheric} but still lack constraints from consistent modeling yet. We find a robust superadiabatic layer at about 7-to-4 bars near the $\rm H_{2}S$ weather layer and 1 bar level near the $\rm CH_{4}$ weather layer (Fig~\ref{Fig1-T-P}\textbf{b} and \textbf{c}). If the deep heavy element abundance is about 50 times solar, the simulated temperature gradient is about -1.2 $\rm K\;km^{-1}$ at $\sim$8 bars, while the local adiabatic lapse rate is about -1.1 $\rm K\;km^{-1}$ (see calculation methods in the caption of Fig~\ref{Fig1-T-P}). At about the 1 bar level, the simulated temperature gradient is about -1.55 $\rm K\;km^{-1}$, while the local adiabatic lapse rate should be only -1 $\rm K\;km^{-1}$. Below 10 bars level, temperature profiles are close to the local adiabat. The superadiabaticity is generally smaller if we reduce the heavy element abundance, but the temperature profiles are still superadiabatic even if the heavy element abundance is only 10 times solar.

Previous 1D theories \citep{guillot1995condensation,li2015saturn,leconte2017condensation,markham2022convective} cannot explain the superadiabatic layer found in our simulations when heavy element abundance is only about 10 times solar. The previous theory predicts that convective inhibition and the subsequent superadiabat only occur if the mass mixing ratio of $\rm CH_{4}$ vapor exceeds 0.10 kg/kg \citep{guillot1995condensation,leconte2017condensation}, which is slightly less than 30 times solar \citep[note that we use different solar abundance from][]{guillot1995condensation,leconte2017condensation}. Using previous theories, one can also estimate the inhibition threshold of $\rm H_{2}S$. The inhibition threshold of $\rm H_{2}S$ is 0.097 kg/kg in a fully saturated weather layer, which is much more than 100 times solar. However, numerical solutions show that there are superadiabats at $\rm CH_{4}$ and $\rm H_{2}S$ weather layers with only 10 times solar (Fig~\ref{Fig1-T-P}\textbf{c}).

\begin{figure*}[t]
\centering
\includegraphics[width=\linewidth]{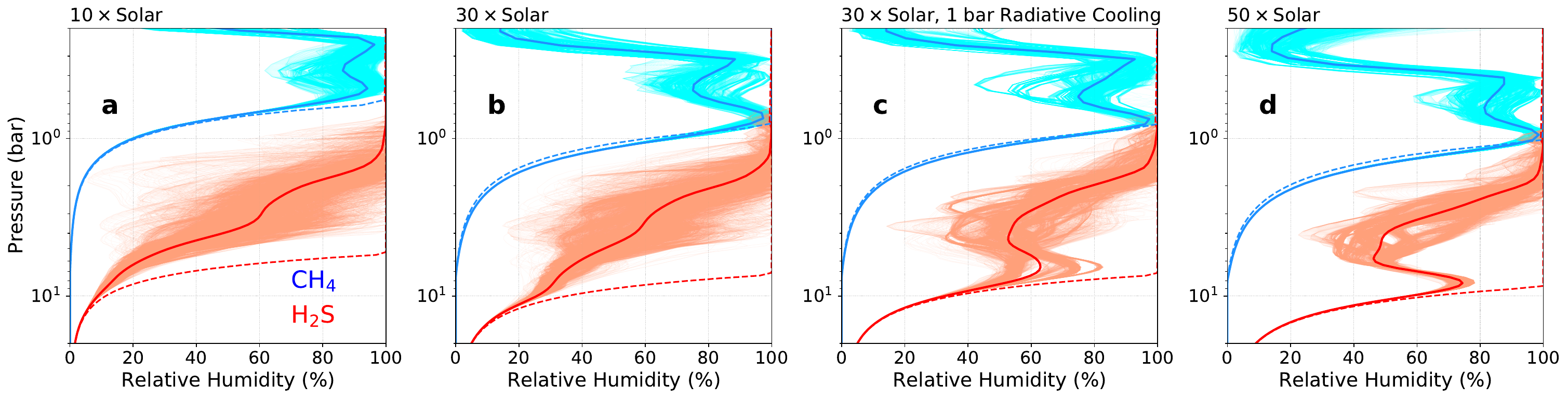}
\caption{Same as Fig~\ref{Fig2-abundance} but for relative humidity. Dashed lines denote the relative humidity at the initial chemical equilibrium state \citep{li2018moist}, and light lines show the temporal and spatial variation of relative humidity. Similar to the snapshots in Fig~\ref{Fig2-abundance}, the output does not capture every timestep. Relative humidity of $\rm CH_{4}$ is slightly larger than the initial condition below the initial lifting condensation level because the superadiabat induced by the condensation of $\rm H_{2}S$ reduces the temperature in that region. Condensation only occurs when the localized air parcel is saturated and the relative humidity is at least 100\%. Snapshots of relative humidity show that $\rm H_{2}S$ is rarely saturated below the 5 bar level for all cases.
}
\label{Fig3-rh}
\end{figure*}

The simulated superadiabatic weather layers are maintained by subsaturated stable layers. Previous linear perturbation analyses usually assume the weather layer is fully saturated and do not include a consistent consideration of mixing induced by net convection with downdrafts \citep{guillot1995condensation,leconte2017condensation,markham2022convective}. Downdrafts can bring dry air downward, mix with the saturated air, and reduce the relative humidity in the weather layer. Therefore, the weather layer could be subsaturated instead of fully saturated. The convection-resolving hydrodynamic code in this study, which differs from 1D linear perturbation theories, can capture the relative humidity change due to the mixing by directly solving the mass continuity equations of each species. 

Simulation results support the idea that weather layers could be subsaturated. Fig~\ref{Fig3-rh} shows that the $\rm CH_{4}$ and $\rm H_{2}S$ weather layers are rarely saturated at the original lifting condensation level (i.e., 9-to-7 bar for $\rm H_{2}S$ and $\rm \sim$ 1 bar for $\rm CH_{4}$ depends on the abundance of these species). If we carefully check the relative humidity in the superadiabatic layer (Fig~\ref{Fig3-rh}), we can find that the averaged relative humidity is about 90\% for 10 times solar of $\rm CH_{4}$ and more than 95\% for 30 times solar or more $\rm CH_{4}$ near the lifting condensation level at the initial state. 

If we apply the convective inhibition threshold for \textit{subsaturated} weather layers derived in Ge \textit{submitted} to $\rm H_{2}S$, the inhibition threshold is greatly reduced. The threshold for the subsaturated atmosphere in Ge \textit{submitted} is,

\begin{equation}
\label{eq:final-eq}
\begin{split}
    q_{\rm inh} & \sim 10 \Big[ \frac{(\gamma-1)^2\sigma^{2}\;T_{\rm eff}^{7}}{\gamma^{2} R_{d} \; p^{2}} \frac{p_0^{R_{d}/c_{p}}}{p^{R_{d}/c_{p}}}\Big]^{1/3} \\
    & \sim 5.252\times 10^{-5} \; T_{\rm eff}^{7/3} \; p^{-(2+R_{d}/c_{p})/3}.
\end{split}
\end{equation}
where $\gamma$ is the polytropic index of the hydrogen-helium mixture; $\sigma$ is the Stephan-Boltzmann constant; $R_{d}$ is the specific gas constant of the hydrogen-helium mixture; $p_{0}$ is the reference pressure at 0.1 bar; $p$ is local atmospheric pressure; and $T_{\rm eff}$ is the effective temperature that $\sigma T_{\rm eff}^4 \sim F_{\rm rad}^{\rm TOA}$. The threshold value of convective inhibition is only about 10 ppm for $\rm H_{2}S$, four orders of magnitude smaller than the one acquired from the previous theory, which assumes the weather layers are fully saturated. This result could explain the superadiabat found at the $\rm H_{2}S$ weather layer and indicate that the saturation condition plays an important role in reducing the threshold of convective inhibition here.

\begin{figure*}[t]
\centering
\includegraphics[width=\linewidth]{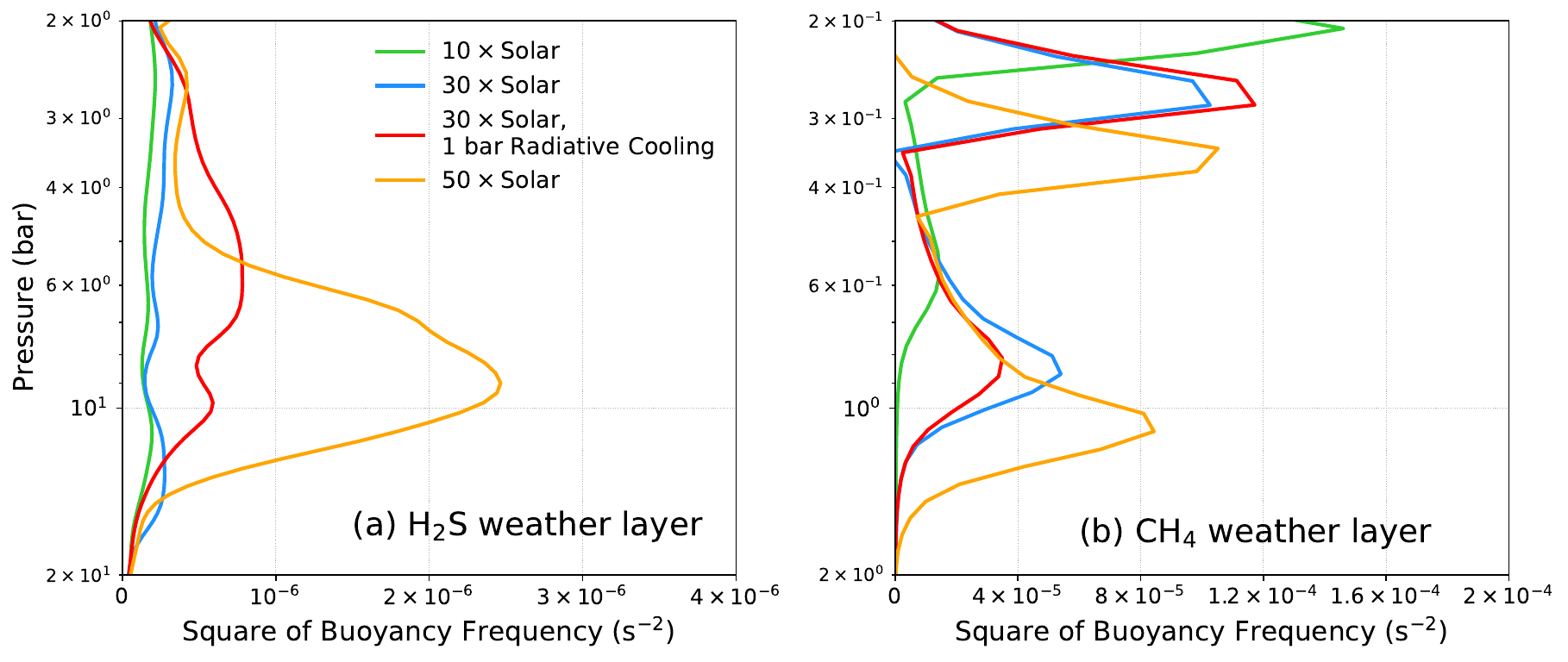}
\caption{The square of the buoyancy frequency at $\rm H_{2}S$ weather layer (\textbf{a}) from 20 bars to 2 bars and $\rm CH_{4}$ weather layer (\textbf{b}) from 2 bars to 0.2 bars. The atmospheric buoyancy frequency is about $\rm 10^{-3} \; s^{-1}$ at about 9 bars level and $\rm 10^{-2} \; s^{-1}$ near the 1 bar level for the case with 50 times solar of $\rm CH_{4}$ and $\rm H_{2}S$.}
\label{Fig4-N-sqr}
\end{figure*}

We can use the virtual potential temperature gradient to estimate the buoyancy frequency of the subsaturated atmosphere (Fig~\ref{Fig4-N-sqr}),

\begin{equation}
    N^{2} \sim \frac{g}{\theta_{v}}\frac{d\theta_v}{dz},
\end{equation}
where $\theta_{v}$ is the virtual potential temperature. The buoyancy frequency, $N$, is about $\rm 10^{-3}\;s^{-1}$ near the $\rm H_{2}S$ weather layer and $\rm 10^{-2}\;s^{-1}$ near the $\rm CH_{4}$ weather layer at about 1 bar for the case with 50 times solar abundance. 

Stable layers prohibit the convective heat flux from passing through them, leading to the net cooling of the atmosphere above stable layers. The net cooling effect induces a colder region than the adiabatic atmosphere above the stable layer and leads to a superadiabat inside and above the stable layer (Fig~\ref{Fig1-T-P}\textbf{c}). The presence of stable layers also indicates that the predominant heat transport mechanism is the latent heat flux delivered by condensed-phase precipitations instead of convective heat flux.

We also find that the thermal structure and buoyancy frequency are slightly sensitive to the thickness of the radiative cooling zone. For nominal cases, in which the radiative cooling extends down to 2 bars (yellow, blue, and green lines in Fig~\ref{Fig1-T-P}), a portion of the net heat flow may pass through the stably stratified layer induced by the $\rm CH_{4}$ weather layer at about 1 bar. As a consequence, the lapse rate of the case with the shallow radiative cooling layer (the red line in Fig~\ref{Fig1-T-P}\textbf{c}) is greater than the one with a deeper radiative cooling layer (the blue line in Fig~\ref{Fig1-T-P}\textbf{c}). This is likely due to the stronger net cooling in the case with the 1-bar-deep radiative layer.

\subsection{Simulated $\rm CH_{4}$ and $\rm H_{2}S$ Cloud Structure}
\label{sec:simulated-cloud}

\begin{figure*}[t]
\centering
\includegraphics[width=\linewidth]{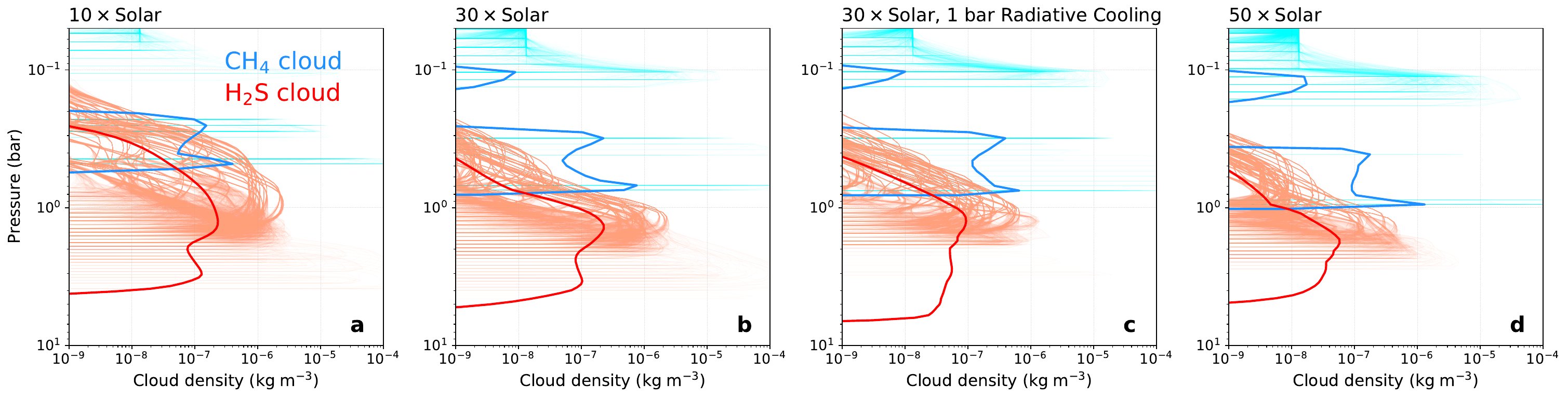}
\caption{The cloud density of the averaged $\rm CH_{4}$ and $\rm H_{2}S$ cloud content (thick lines) and intermittent localized storms (thin lines). Thinner lines denote localized and transient cloud content in each vertical column. The $\rm CH_{4}$ cloud density inside the storm could be more than $\rm 10^{-4}$-to-$\rm 10^{-3}\; kg \; m^{-3}$ but these storms cannot last long and are eventually turned off by the exceedingly large latent heat flux compared with the averaged planetary heat flux. It is also noticeable that this dataset does not capture the variation of storms at each timestep.}
\label{Fig5-cloud}
\end{figure*}

Fig~\ref{Fig5-cloud} shows the averaged (thick lines) and temporal variation (light lines) of cloud density. The temporal variation of localized clouds denotes there are small-scale, intermittent, and localized storms. We first find that the averaged cloud density and the overall vertical cloud structure are not sensitive to the metallicity of heavy elements. Secondly, there are three distinct cloud layers in the case with 50 times solar, which might be close to the real composition of Neptune's atmosphere: (1) $\rm H_{2}S$ clouds reside at about 5-to-2 bars; (2) a major $\rm CH_{4}$ cloud layer with a thin but abundant cloud base near $\sim$ 1 bar and a less abundant cloud layer extended up to about 0.4 bars; (3) a detached and even less abundant $\rm CH_{4}$ cloud layer near 0.1 bar. Three cloud layers are detached by two stably stratified layers at about 1 bar and 0.4-to-0.3 bars (Fig~\ref{Fig4-N-sqr}\textbf{b}). The simulated cloud layers' location and structure are similar to Neptune's aerosol structure retrieved from recent visible/near-IR observations \citep{irwin2022hazy}. However, we do not find a significant amount of clouds within stably stratified layers. Another discrepancy might be the composition of aerosol near 1 bar level. \cite{irwin2022hazy} found the optical properties of these aerosol particles are more like to be photochemical haze instead of submicron cloud particles.

The cloud density varies between $10^{-7}-10^{-6}$ $\rm kg\;m^{-3}$ in altitude. This simulated number is close to the upper limit estimated by Eq~\ref{eq:cloud-limit} and significantly lower than the predicted value from previous 1D equilibrium cloud condensation models (ECCM) \citep[e.g.,][]{weidenschilling1973atmospheric,hueso2020convective}. ECCM suggests that the maximum density of cloud content is about $10^{-2}$-to-1 $\rm kg\;m^{-3}$ and sensitive to the abundance of cloud-forming species in the atmosphere. A disparity between ECCM and our model is that ECCM does not consider precipitations and their evaporation and omits the physics of latent heat flux. The abundant clouds suggested by ECCM would eventually form precipitating condensates after a timescale that is longer than the storm itself \citep{carlson1987cloud} and lead to a latent heat flux that greatly exceeds the planetary heat flux and turn off convection. Therefore, from a global view, omitting precipitations and heat delivered by latent heating can lead to an overestimated cloud content in weakly forced giant planets.

Light lines in Fig~\ref{Fig5-cloud} indicate there are short-lived and intense storms with abundant fresh clouds, similar to results from previous numerical studies about Jupiter's weather cycles \citep[e.g.,][]{sugiyama2011intermittent,sugiyama2014numerical}. The instant cloud density could exceed $10^{-4} \; {\rm kg\;m^{-3}}$, which is 100 times more than the average value. Most clouds and precipitations are formed during these intermittent storms in the $\rm CH_{4}$ weather layer. The $\rm CH_{4}$ cloud layer is generally more intermittent and stormy with more variations than the $\rm H_{2}S$ cloud layer.

The variation of cloud density due to the occurrence of storms complicates the cloud microphysics and the choice of microphysics parameters in simulations since these parameters are sensitive to the localized and transient cloud density (i.e., collision rate of cloud particles) instead of the disk-averaged values. Therefore, we choose microphysics parameters, such as settling velocity and autoconversion timescale, that can represent the cloud microphysics inside the storm instead of directly determining the cloud lifetime from the estimated cloud density (see Section~\ref{sec:cloud} and \ref{sec:methods}). Future studies about cloud microphysics and the origin of intermittent storms in hydrogen atmospheres are necessary to help us fully understand the cloud formation in ice giants and perhaps other giant planet atmospheres.

\subsection{Simulated Eddy Diffusivity of \texorpdfstring{$\rm CH_{4}$}{Lg} and \texorpdfstring{$\rm H_{2}S$}{Lg}}
\label{sec:simulated-kzz}

\begin{figure*}[t]
\centering
\includegraphics[width=\linewidth]{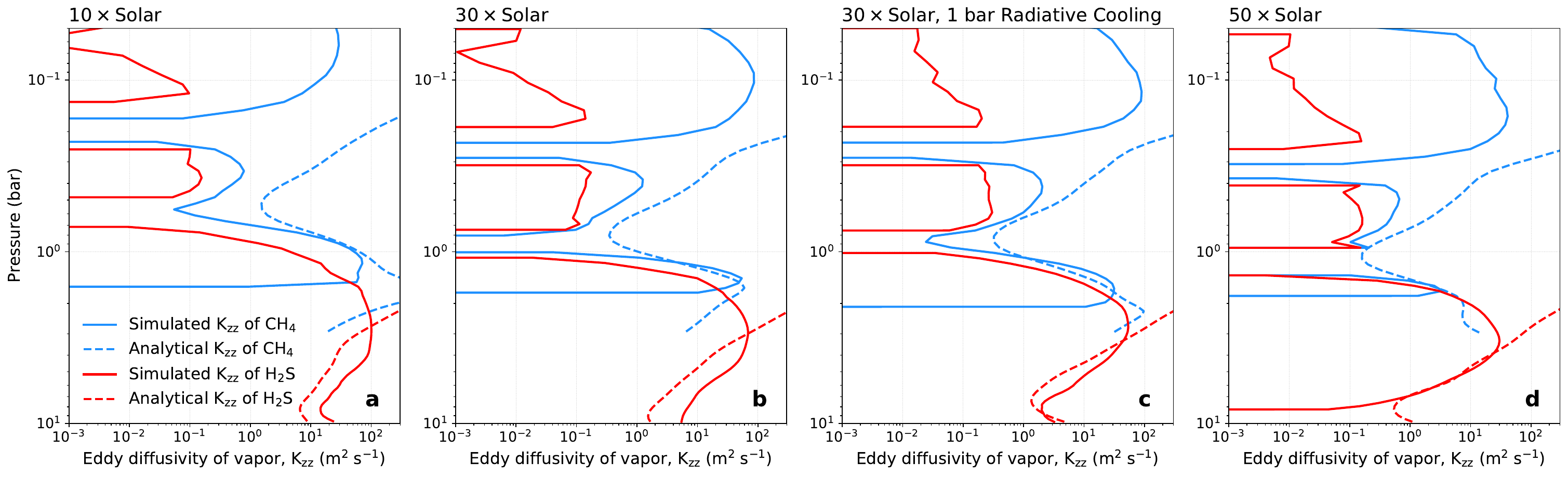}
\caption{The averaged eddy diffusivity of $\rm CH_{4}$ (blue lines) and $\rm H_{2}S$ (red lines) vapor from numerical solutions (solid lines) and analytical solutions (dashed lines). The analytical eddy diffusivity is calculated by the simulated concentration gradient using Eq~\ref{eq:Kzz-general}. In most layers, the analytical eddy diffusivity (dashed lines) is larger than or very close to the numerical solutions of the same species (solid lines). Not the analytical $K_{zz}$ in this figure is smaller than the estimated numbers in Section~\ref{sec:kzz} because of the superadiabatic and subsaturated weather layers (see Appendix~\ref{app-kzz} for the derivation).}
\label{Fig6-kzz}
\end{figure*}

Different from the cloud density, which is not sensitive to the abundance of cloud-forming species, simulation results show that the eddy diffusivity tends to be smaller with higher metallicity (Fig~\ref{Fig6-kzz}). This is consistent with the conclusion from the analytical analysis. The simulated eddy diffusivity (solid lines in Fig~\ref{Fig6-kzz}) is close to the upper limit estimated by Eq~\ref{eq:Kzz-general} (dashed lines in Fig~\ref{Fig6-kzz}). The eddy diffusivity is about $\rm 30 \; m^{2}\;s^{-1}$ at the $\rm H_{2}S$ cloud level and $\rm 1 \; m^{2}\;s^{-1}$ at the $\rm CH_{4}$ cloud level for 50 times solar of cloud-forming species. Eddy diffusivity at the stably stratified layer is significantly smaller than the number estimated from mixing length theory \citep[e.g.,][]{wang2015new,zhang2020atmospheric,moses2020atmospheric}, as analytical estimations suggested in Section~\ref{sec:kzz}.

Simulation results confirm that moist convection and mixing efficiency are strictly limited by the heat flow and show the analytically estimated eddy diffusivity of moisture (Section~\ref{sec:self-regulation}) is valid. Because thermal convection is inhibited at the stratified layer, heat is primarily delivered across the stable layer through precipitations and latent heating, in which $F_{\rm latent} \sim F_{\rm rad}^{\rm TOA}$. Therefore, the simulated cloud density and eddy diffusivities are close to the upper limits estimated in Section~\ref{sec:self-regulation}.

\section{Discussions and Implications}\label{sec:discussions}

\subsection{Does Radiation Transport Heat at the $CH_{4}$ Weather Layer}\label{sec:discuss-rad}

A superadiabatic layer may open a channel for radiative heat transport since the radiative heat flux is proportional to the temperature gradient according to the approximation of radiative diffusion. Radiation should be more important in the $\rm CH_{4}$ weather layer than in the $\rm H_{2}S$ layer since the atmosphere is generally less opaque in low-pressure regions. Here, we employ Rossland mean opacity from \cite{freedman2014gaseous} at 1 bar and 80 K ($\kappa\sim 10^{-4}$ $\rm kg^{-1}\;m^{2}$) to estimate the radiative diffusion timescale at the $\sim$1 bar $\rm CH_{4}$ weather layer \citep{zhang2020atmospheric},

\begin{equation}
\begin{split}
    \tau_{\rm rad} & \sim \frac{p^{2}c_{p}\kappa}{g^{2}\sigma T^{3}} 
                   \sim \frac{(10^5)^2\times 10^4 \times 10^{-4}}{10^2\times 5\times 10^{-8} \times 80^3} \\
                   & \sim 3\times 10^9 \; {\rm s} \sim 100 \; {\rm Earth\;Year}.
\end{split}
\end{equation}
On the other hand, the sedimentation timescale of $\rm CH_{4}$ precipitation, which determines how fast latent heat delivers energy across the weather layer, should be shorter than

\begin{equation}
    \tau \sim \frac{H}{V_{T}} < \frac{3\times 10^{4}\; {\rm m}}{0.1 \; {\rm m\;s^{-1}}} \sim 3 \; {\rm days}
\end{equation}
with a very slow sedimentation speed, it is clearly shorter than the radiative cooling timescale at the 1 bar level. Hence, ignoring the radiative cooling in the major $\rm CH_{4}$ weather layer ($\sim$1 bar) might be valid. 

However, the radiative timescale of the secondary $\rm CH_{4}$ weather layer near the tropopause ($\sim$0.1 bar) is reduced to

\begin{equation}
\begin{split}
    \tau_{\rm rad} & \sim \frac{(10^4)^2\times 10^4 \times 10^{-4}}{10^2\times 5\times 10^{-8} \times 50^3} \\
                   & \sim 10^7 \; {\rm s} \sim 0.3 \; {\rm Earth\;Year}.
\end{split}
\end{equation}
This number suggests that radiation might play an important role in heat transport and cloud formation near the tropopause. Besides, heating due to hydrocarbon absorption would alter the temperature structure to a subadiabat. After all, although there is a $\rm CH_{4}$ cloud layer near 0.1 bar in our simulation, we are not certain whether it would exist if we included a more realistic radiative transfer scheme to simulate the temperature profile and mixing efficiency near the tropopause. Simulation in this research also cannot explain the observed difference in the cloud structure between two ice giants \citep{irwin2022hazy}.

\subsection{Cloud Opacity and the Possible Difference between Uranus and Neptune}\label{sec:cloud-opacity}

If cloud content is more depleted than previously expected in the weather layer, as proposed in this paper, the cloud opacity should be significantly reduced. One may wonder whether the cloudless weather layer can provide enough opacity to match the recent observation \citep{irwin2022hazy}. The cloud opacity shown in the later part of this study has the same definition as the one in \cite{irwin2022hazy} but is sometimes referred to as the cloud optical depth in other articles.

Similar to the density and column density of clouds, we can estimate a range of cloud opacity from the latent heat flux. We assume the cloud particles have the same size and they are spherical. The cloud column number density in $\rm m^{-2}$, $N_cH_c$, is determined by the column density and the mass of each particle with diameter $D$,

\begin{equation}
    N_cH_c \sim \dfrac{\sigma_{c}}{\dfrac{\pi}{6}\rho_{l} D^3},
\end{equation}
where $\rho_{l}$ is the density of cloud particles (liquid or solid ice). The optical depth can be calculated by the column density multiplied by the extinction cross-section, $\sigma_{\rm ext}\pi D^{2}/4$, of each cloud particle,

\begin{equation}
    \tau \sim \frac{\pi}{4}\sigma_{\rm ext}D^2 N_cH_c \sim \dfrac{3\rho_{c}H_{c}\sigma_{\rm ext}}{2\rho_{l} D} < \frac{F_{\rm rad}^{\rm TOA}\tau_{\rm c \rightarrow p}\sigma_{\rm ext}}{ \rho_{l} L_{v} D},
\end{equation}

If we plug in some typical numbers of ice giant atmospheres into these equations, we obtain the optical depth, $\tau$, as a function of autoconversion timescale, particle diameter, and the optical extinction coefficient,

\begin{equation}
\begin{split}
    \tau & \sim \dfrac{0.5 {\rm \;W\;m^{-2}} }{5\times 10^{5} {\rm \;J\;kg^{-1}} \times 10^{3} {\rm \;kg\;m^{-3}}}\dfrac{\tau_{\rm c \rightarrow p}\sigma_{\rm ext}}{D} \\
         & \sim 10^{-9} \dfrac{\tau_{\rm c \rightarrow p}\sigma_{\rm ext}}{D}.
\end{split}
\end{equation}
We then assume the extinction coefficient $\sigma_{\rm ext}$ is on the order of unity. If the particle diameter is about 10 $\rm \mu m$, and the autoconversion timescale is about $10^{4}$ s, then the globally averaged cloud optical depth is about 1. 

Alternatively, if we use the aerosol particle size (i.e., diameter) suggested by observations \citep{irwin2022hazy}, which is around 1 $\rm \mu m$, then the upper bound of cloud optical depth is about 10. The observed aerosol optical depth of Neptune is about 0.81 for the layer at $\sim$5 bars, 1.8 for the layer at $\sim$2 bars, and 0.05 for the layer near the tropopause \citep[Section 3.9 in][]{irwin2022hazy}, which roughly fit into the range estimated by our analytical analysis in the order of magnitude. The simulated cloud structure also has a more abundant cloud layer at the 1-to-0.4 bars $\rm CH_{4}$ weather layer than $\rm H_{2}S$ cloud layer and the $\rm CH_{4}$ cloud layer near the tropopause, which is similar to the vertical opacity variations retrieved from observations. On the other hand, the integrated optical depth estimated from the ECCM should be much larger than the observed value with much more abundant clouds. The better congruence of our model with the observational data strongly implies that the regulatory mechanism of moist convection on ice giants is likely self-limited by the planetary heat flux.

However, it is also worth noticing that the concept of cloud in this study does not include photochemical haze and chromophores, which universally exist on both ice giants and might have a significant contribution to the observed cloud opacity. They usually have a longer lifetime than cloud particles, and small haze particles may serve as cloud condensation nuclei (CCN). But they are not directly formed by phase transitions like clouds, and, therefore, perhaps they have limited contribution to heat transport but are not trivial for the observed aerosol opacity. A notable difference between our theory and simulations with the retrieved aerosol structure in \cite{irwin2022hazy} is that we do not find clouds in stably stratified layers. On the flip side, the observation suggests there should be aerosols. The aerosol particles might be haze particles trapped in the stable layers. Cloud particles in different pressure levels are also not necessary to have the same shape, size, or optical properties (i.e., extinction coefficient) as we assumed in this section. The purpose of estimating the cloud opacity is merely to ensure that the reduced cloud density suggested by heat-flux limited cloud formation can provide enough opacity compared with observations. 

Our calculation further implies that there should be less cloud on Uranus than on Neptune due to Uranus' smaller heat flux \citep{pollack1986estimates} if the overall cloud lifetime is similar for both planets. Thus, the observed Uranus haze layer at the 1–2-bar \citep[Aerosol-2 layer in][]{irwin2022hazy}, which is thicker than the one on Neptune, might suggest an important contribution of the photochemical haze production in Uranus's atmosphere. On the other hand, it is also possible that Uranus' clouds are less precipitable with a longer autoconversion timescale than Neptune's since Uranus' atmosphere is more tranquil and less stormy. Future cloud microphysics studies about $\rm CH_{4}$ and $\rm H_{2}S$ clouds can look into this issue. It is also possible that Voyager 2 captured a particular and relatively quiet period of Uranus with weaker-than-average heat flux. A future orbiter with long-term monitoring can better constrain Uranus' heat flux with long-term monitoring.

\subsection{Implication for Radio Observations}
\label{sec:implication-radio}

The brightness temperature obtained from microwave/radio observations provides valuable information on the temperature and compositional structure beneath the opaque cloud layers down to tens of bars. However, these observations face a challenge of a degeneracy in which the contributions from gas opacity and kinetic temperature are difficult to distinguish \citep[e.g.,][]{hofstadter1990vertical,li2017distribution,de2019jupiter,tollefson2019neptune,tollefson2021neptune,molter2021tropospheric,moeckel2023ammonia,akins2023evidence}. 

Observers usually employ dry or moist adiabats in radiative transfer models to retrieve the distribution of opacity sources as a function of pressure and latitude \citep{li2017distribution,de2019first,tollefson2021neptune}. However, radio retrievals may need to consider the possibility of superadiabatic temperature profiles at the condensation level and slightly subadiabatic profiles in the free-convection zone since the mass-loading effect may have a broader impact on temperature than previously recognized, as suggested in Section~\ref{sec:T-N^2}. If temperature profiles are superadiabatic at the condensation level, it may suggest that the increased emission due to the superadiabat is currently interpreted as $\rm H_{2}S$ depletion in ice giants \citep{tollefson2019neptune,molter2021tropospheric,tollefson2021neptune}. Since this study also suggests that cloud content is less cloudy than expected, radio observers may also not need to worry about the scattering due to large precipitating particles. 

Here, we propose that numerical simulations can provide alternative and physically reasonable temperature profiles to aid observers in retrieving the distribution of opacity sources. In the future, forward modeling with both the simulated temperature structure and opacity sources may be necessary to explain the brightness temperature and provide insight into the role of moist convection in the atmospheres of ice giants and other giant planets. This modeling should involve more complex physics that incorporates moist convection and large-scale atmospheric dynamics, which could help improve retrievals' accuracy and enhance our understanding of the atmosphere's behavior.

\subsection{Implication for Uranus Orbiter and Probe}
\label{sec:implication-UOP}

Ice giants also provide a unique regime to understand moist convection in giant planets \citep{hofstadter2020future,guillot2022uranus} and an opportunity to test the self-limited moist convection and convective inhibition proposed in this study. The vertical temperature and composition structure at the $\rm CH_{4}$ condensation level can be constrained by radio occultation from the orbiter \citep{lindal1987atmosphere,lindal1992atmosphere}. However, the previous glimpse only provided information about a specific location. The upcoming Uranus orbiter will have the potential to map the atmosphere by radio occultations at different latitudes and inside-outside storms, which can provide a global view of the thermal and compositional structure. 

In-situ measurements from the probe and remote sensing can further explore whether the superadiabatic temperature structure can be sustained in $\rm CH_{4}$ and $\rm H_{2}S$ weather layers and whether the stably stratified layers are subsaturated. Both layers are thinner than one scale height, necessitating the dropped-in probe to be capable of collecting at least ten samples within a 10 km distance to resolve the compositional and temperature gradient at the weather layer. Although the Galileo probe observed a tentative superadiabatic temperature structure at around 10 bars level in Jupiter \citep{magalhaes2002stratification}, the sampling frequency of water abundance was insufficient to determine whether the observed superadiabat was caused by the mass-loading effect \citep{wong2004updated}. The scientific data obtained from the probe can significantly improve our understanding of how moist convection and cloud formation affect general circulation in hydrogen atmospheres \citep{fletcher2020ice}.

\subsection{Implication for the other Giant Planets}
\label{sec:implication-giant-planets}

Perhaps one of the most important implications of this study to the atmospheric dynamics in the other giant planets is that the convective mixing has a much longer eddy mixing timescale than previously estimated, with a significantly reduced eddy diffusivity. The efficient heat transport due to the latent heating and cooling leads to less material exchange in giant planet weather layers. 

Using Saturn as an example, if the vertical heat transport is solely controlled by convective heat flux (i.e., the buoyancy of air parcels), one can use the mixing length theory \citep[MLT, e.g.,][]{showman2013atmospheric,wang2015new,zhang2020atmospheric} to estimate the eddy diffusivity, $K_{zz} \sim wH \sim 10^{5} \; {\rm m^{2}\;s^{-1}}$, in Saturn's troposphere. However, if we take the latent heat flux due to the cloud formation of water into account, the $K_{zz, {\rm H_{2}O}}$ is then reduced to the level of 0.1-to-1 $\rm m^{2}\;s^{-1}$ with 10 times solar of water. The $K_{zz, {\rm H_{2}O}}$ may vary in one order of magnitude because it depends on the background temperature and relative humidity (see Appendix~\ref{app-kzz}) since the mass-loading effect may change the thermal profile to superadiabatic \citep{guillot1995condensation,li2015moist,leconte2017condensation,markham2022convective} by assuming heat is completely delivered by buoyancy-driven convection. If we choose the pressure scale height at the water condensation level, $H \sim 10^{5} \; {\rm m}$, as the length scale for the eddy diffusion process, the eddy diffusive timescale is about $\tau \sim H^2/{K_{zz, {\rm H_{2}O}}} \gtrsim 10^{10} \; {\rm s} \sim 300 \; {\rm yrs}$. The assumption that the thickness of the low mixing region is approximately one scale height is supported by ice giant simulations (Fig~\ref{Fig6-kzz}), where the simulated $K_{zz}$ of $\rm CH_{4}$ is smaller than 1 $\rm m^{2}\;s^{-1}$ from 1 bar to 0.4 bars. Whether this is true on Saturn requires future numerical simulations to confirm. A recent radio observation found remnants of giant storms that happened on Saturn a hundred years ago \citep{li2023long}. Remnants of storms cannot last for a century in a fully convective troposphere since convection would quickly smear out remnants. However, the reduced mixing efficiency proposed in this study might be able to explain this observation. However, the choice of the mixing length scale for the 1D eddy diffusion process needs more support from simulations. Future large-scale simulations \citep[for example,][]{lian2010generation,liu2010mechanisms,young2019simulatingI,young2019simulatingII,sankar2022new} are necessary to explore the hydrological timescale as a function of metallicity, heat flux, and gravity to understand the longer-than-expected convective timescales in giant planet tropospheres.

\begin{table*}[t]
  \centering
  \begin{tabular}{l|c|c|c|c|c}
    \toprule
    Substellar Objects & Neptune & Saturn & Jupiter & Type-Y Dwarf & Type-T Dwarf \\
    \midrule
    (A) Effective Temperature & $\sim$ 50 K & $\sim$ 95 K & $\sim$ 105 K & $\sim$ 300 K & $\sim$ 800 K \\
    (B) Cloud Species & $\rm CH_{4}$ & $\rm H_{2}O$ & $\rm H_{2}O$ & $\rm H_{2}O$ & $\rm MgSiO_{3}$ \\
    (C) Metallicity [times solar] & 80 & 10 & 3 & 1 & 1 \\
    (D) Weather Layer Atmospheric Pressure [bars] & $\sim$ 1 & $\sim$ 20 $^{[1]}$ & $\sim$ 7 & $\sim$ 0.3 $^{[2]}$ & $\sim$ 50 $^{[3]}$ \\
    (E) Weather Layer Temperature [K] & $\sim$ 80 & $\sim$ 350 $^{[1]}$ & $\sim$ 300 & $\sim$ 270 $^{[2]}$ & $\sim$ 1800 $^{[3]}$ \\
    (F) Gravity [$\rm m\;s^{-2}$] & 11.2 & 10.4 & 24.8 & $\sim 10^2$ $^{[4]}$ & $\sim 3\times 10^2$ $^{[4]}$ \\
    (G) Local Scale Height [m] & $\rm \sim 3\times 10^4$ & $\rm \sim 10^5$ & $\rm \sim 4\times 10^4$ & $\sim 10^4$ & $\sim 2\times 10^4$ \\
    (H) Dimensionless Latent Heat, $\beta$ & 11 & 24 & 24 & 24 & 7 $^{[5]}$ \\
    \midrule
    (I) MLT $K_{zz}$ [$\rm m^{2}\;s^{-1}$] & $\sim 3\times 10^4$ & $\sim 10^5$ & $\sim 5\times 10^4$ & $\sim 10^5$ & $\sim 10^6$ \\
    (J) Moisture $K_{zz,m}$ [$\rm m^{2}\;s^{-1}$] & $\sim$ 0.3 & $\sim$ 1 & $\sim$ 3 & $\sim 2\times 10^3$ & $\sim 10^5$\\
    (K) Moisture Transport Timescale (Earth unit)& $\sim$ 100 yrs & $\sim$ 300 yrs & $\sim$ 10 yrs & $\sim$ 1 day & $\sim$ 1 hr \\
    \bottomrule
  \end{tabular}
  \caption{Eddy diffusivity and convective timescale of the major weather layer in Neptune, Saturn, Jupiter, 300 K type-Y brown dwarfs, and 800 K type-T brown dwarfs. The moisture transport timescale is calculated by $H^2/K_{zz,m}$. References: [1] Fig 1 in \cite{li2023long}; [2] Fig 5 in \cite{lacy2023self}; [3] Fig 1 in \cite{lefevre2022cloud}; [4] Fig 4 in \cite{lacy2023self} (note the unit of gravity in their study is $\rm cm\;s^{-2}$); [5] Eq 20 in \cite{visscher2010atmospheric}.
  }
  \label{tab:HR-diagram}
\end{table*}

We can also explore the timescale of moist convection in giant planet weather layers from the aspect of their planetary heat flux/luminosity/effective temperature. Here, we list the eddy diffusivity of moisture and the convective timescale of moist convection in the major weather layers of substellar objects from the cold ice giants to Jupiter to warm free-floating type T brown dwarfs with different condensates (Table~\ref{tab:HR-diagram}). Major cloud-forming species refer to the most abundant and observable moisture that can condense in the atmosphere. As previous sections discussed, the major condensates are $\rm CH_{4}$ on ice giants. Although $\rm H_{2}O$ may also condense in ice giants with a significant vapor abundance, they are too deep to constrain their behaviors in the foreseeable future, so $\rm H_{2}O$ in ice giants is not discussed here. Due to the same reason, we do not show the estimation of $\rm MgSiO_{3}$ in Jupiter \citep[e.g.,][]{markham2018excitation,zhang2020atmospheric} and Y dwarfs. One can estimate them using Eq~\ref{eq:Kzz-final}. On Saturn, Jupiter, and cold Y dwarfs \citep[e.g.,][]{morley2014water,leggett2015near,lacy2023self}, the major condensates are $\rm H_{2}O$. We choose $\rm MgSiO_{3}$ as the major condensates for late T dwarfs since Mg and Si are the most abundant elements that can form condensates in the warm atmosphere of T dwarfs \citep[e.g.,][]{gao2020aerosol}. 

A short conclusion is that we find, even for 800 K type-T dwarfs, which have significant luminosity, the moisture $K_{zz}$, which includes latent heat flux, is ten times smaller than the number obtained from the mixing length theory by assuming the heat is delivered by buoyancy-driven convection if the mixing length scale is one scale height (Table~\ref{tab:HR-diagram}). The difference between the two mixing efficiency indicates that it is essential to include the physics of latent heat flux and moist convection with complete hydrological cycles on ``cold'' giant planets ($T_{\rm eff} <$ 800 K).

We also emphasize that the theory proposed in this study can only be applied to giant planets far from their host stars or free-floating giant planets (e.g., field brown dwarfs) but not to strongly irradiated giant planets, such as hot Jupiter, where the observable atmospheric motion is dictated by the radiation from host stars instead of their cooling.

\section{Conclusions}
\label{sec:conclusion}

In this study, starting from the first-principle physics of heat transport and continuity equations, we utilize analytical analysis to constrain the cloud density and mixing efficiency of cloud-forming species in giant planet atmospheres. Then, we apply this theory to study ice giants. We suggest that heat transport on weakly forced giant planets limits the cloud formation and mixing efficiency of moisture associated with moist convection. We employ the nonhydrostatic model, SNAP, to explore the temperature and cloud structure in Neptune's atmosphere and validate our analytical theories in Neptune's regime. We show that the simulated cloud density and eddy diffusivity are consistent with analytical solutions and similar to the findings from the recent observations \citep{irwin2022hazy}. We list the main takeaway points here:

\begin{enumerate}
    \item Moist convection is limited by the planetary heat flux in ice giants and possibly in the other weakly forced giant planets because phase transition associated with the cloud formation can efficiently deliver heat across the atmosphere through a complete hydrological cycle of cloud-forming species. Our major findings are that weakly forced giant planets are less cloudy than previously expected (Section~\ref{sec:cloud} \& \ref{sec:simulated-cloud}), the weather layer mixing is less efficient than previously expected (Section~\ref{sec:kzz} \& \ref{sec:simulated-kzz}), and the convective timescale in weather layers is longer than previously expected (Section~\ref{sec:implication-giant-planets}).
    
    \item The upper limit of the globally averaged cloud column density is independent of the atmospheric composition but determined by the planetary heat flux and cloud lifetime,
    
    \begin{equation}
        \sigma_{c} < \frac{F_{\rm rad}^{\rm TOA}\tau_{\rm c \rightarrow p}}{L_{v}}.
    \end{equation}
    
    By assuming the cloud density is uniformly distributed in altitude, we get the upper limit of the cloud density, 
    
    \begin{equation}
        \rho_{c} < \frac{F_{\rm rad}^{\rm TOA}\tau_{\rm c \rightarrow p}}{L_{v}H_{c}}.
    \end{equation}
    
    The upper limit of cloud density is irrelevant to most microphysics processes but is only determined by the autoconversion timescale. The upper limits of $\rm CH_{4}$ and $\rm H_{2}S$ cloud density on Neptune are about $10^{-6}$ $\rm kg\;m^{-3}$ using the observed planetary heat flux from Voyager 2 and assuming the conversion timescale from cloud particles to precipitations is about $\rm 10^4$ s. This estimated cloud density is about 4-to-5 orders of magnitude smaller than the previous ECCM suggested \citep{weidenschilling1973atmospheric,hueso2020convective,atreya2020deep}. We suggest ECCM is more proper to study cloud formation within localized and transient storms without precipitations. In contrast, the theory proposed in this paper is suitable for studying globally averaged cloud contents with complete hydrological cycles.
    
    \item The upper limit of globally averaged eddy diffusivity of moisture is determined by three parameters: the planetary heat flux, atmospheric metallicity, and gravity of the planet. An analytical approximation for the upper limit of eddy diffusivity in an adiabatic and saturated atmosphere is,
    
    \begin{equation}
    \label{eq:Kzz-final}
        K_{zz} < \frac{F_{\rm rad}^{\rm TOA}H}{\rho L_{v} \epsilon \eta_s} \Big( \frac{\gamma-1}{\gamma}\beta-1 \Big)^{-1}. 
    \end{equation}
    The eddy diffusivity of moisture is not directly related to cloud properties. The upper limits of eddy diffusivity of $\rm CH_{4}$ and $\rm H_{2}S$ should be less than $1$ $\rm m^{2}\;s^{-1}$ and $30$ $\rm m^{2}\;s^{-1}$, respectively, at their condensation levels. Simulation suggests the $K_{zz}$ is about 0.2 $\rm m^{2}\;s^{-1}$ at about $\sim$ 1 bar level and 30 $\rm m^{2}\;s^{-1}$ at about $\sim$ 4 bar level. The analytically estimated and simulated eddy diffusivities are all a few orders of magnitude smaller than the previous mixing length theory suggested \citep[e.g.,][]{showman2013atmospheric,wang2015new,zhang2020atmospheric,moses2020atmospheric,irwin2022hazy} by taking the complete hydrological cycle with the latent heat flux into account. In the discussion, we further show that the hydrological timescale of the other giant planets is also longer than previously expected.

    \item Nonhydrostatic simulation shows that the condensation of $\rm CH_{4}$ and $\rm H_{2}S$ induces two stably stratified layers at about 1 bar and 8 bars for cases with deep heavy element abundance ranging from 30 times solar to 50 times solar. The temperature profile is superadiabatic in these stable layers. We find convection is inhibited in this layer, and convective inhibition is induced by the compositional gradient in subsaturated weather layers. 

    \item The simulated cloud structure has three layers. The bottom one extends from about 5-to-2 bars and is composed of $\rm H_{2}S$ clouds. The middle cloud layer is composed of a mixture of $\rm H_{2}S$ and $\rm CH_{4}$ clouds but mostly $\rm CH_{4}$ clouds with an abundant and thin cloud base near 1 bar level and an extended but less cloudy layer to 0.4 bars level. These two layers are similar to the retrieved tropospheric aerosol structure in \cite{irwin2022hazy}. We also find a topmost intermittent $\rm CH_{4}$ cloud layer located at 0.2-to-0.1 bar. However, since we greatly simplify the radiative transfer near the tropopause, future studies with a radiative transfer scheme may need to test whether this topmost cloud layer is robust and whether it is the aerosol layer reported in \cite{irwin2022hazy} and \cite{chavez2023evolution}.

    \item We estimated the cloud optical depth from the upper limit of cloud column density constrained by the heat transport. We show that the total optical depth of clouds is,
    
    \begin{equation}
    \tau < \frac{F_{\rm rad}^{\rm TOA}\tau_{\rm c \rightarrow p}\sigma_{\rm ext}}{ \rho_{l} L_{v} D},
    \end{equation}
    by assuming cloud particles have a uniform size and the same extinction coefficient. The estimated cloud opacity is about 1-to-10 for Neptune. The low cloud density proposed by our theory does not lead to a cloudless atmosphere in optical observations. The estimated cloud opacity is similar to the retrieved aerosol opacity on Neptune \citep{irwin2022hazy}. We further show that if the planetary heat flux of Uranus is near-zero, as suggested by observations from Voyager 2, the production rate of tropospheric haze or chromophores on Uranus should be significantly larger than the one on Neptune. It is also possible that the cloud lifetime in Uranus' tranquil atmosphere is much longer than Neptune's.
    
    \item We provide implications for future radio observations and the Uranus Orbiter and Probe. Radio occultations measured by the orbiter and in-situ measurements from the probe can provide detailed information about the temperature structure, cloud density, and relative humidity to help us better understand the moist convection in the hydrogen atmosphere and validate the theories proposed in this paper. We suggest the sampling frequency of in-situ measurement should be at least ten times in 10 km to resolve the superadiabatic temperature profile inside the weather layer.    
    
\end{enumerate}

Our study steps forward from the picture proposed by \cite{ackerman2001precipitating} by taking latent heat flux into account. $K_{zz}$ and $f_{\rm sed}$ are two dominating parameters in \cite{ackerman2001precipitating}. Our study shows that the upper bound of $K_{zz}$ can be directly estimated by the latent heat flux, and the physics of $f_{\rm sed}$ is related to the cloud lifetime (e.g., autoconversion timescale).

Future work may need to extend the simulation part into 3D domains with more realistic radiative transfer to study the $\rm CH_{4}$ cloud formation near the tropopause. Hydrocarbons have strong radiative heating and cooling effects near the tropopause \citep{li2018high}, and therefore, the mixing efficiency and the cloud content near the tropopause might be reduced compared to the simulated result from this work. Radiative transfer with self-consistent mixing and transport of hydrocarbons is computationally expensive. It might be feasible in the future to incorporate a more realistic cooling and heating profile from radiative transfer models into these hydrodynamic simulations.

It is also interesting to see how the stably stratified layers in weather zones impact the large-scale dynamics. Future large-scale simulations may tell us the answer. Inertia gravity waves formed in the stably stratified layer are usually considered a source of momentum drag, which might be important to jet formation in ice giants. 

This study serves as a pioneer work to understand the cloud activity and moist convection on ice giants and other weakly forced giant plants from first-principle physics and localized cloud-resolving simulations. It provides important implications for future observations and missions like the Uranus Orbiter and Probe to enhance their scientific returns. This study shows that sophisticated cloud-resolving simulations can provide insightful hints for simple analytical analysis to shed light on the complicated atmospheric dynamics in hydrogen atmospheres, which involve heat transport, phase transition, disequilibrium chemistry, and mass-loading effect.

\section*{acknowledgments}
We acknowledge Xiaoshan Huang for making the conceptual art in Fig 1. H.G. acknowledges useful discussions with Micheal H. Wong, Peter Gao, Imke de Pater, Edward Molter, Tristan Guillot, and Andrew P. Ingersoll. We acknowledge the useful and constructive comments from two anonymous referees. H.G. is supported by NASA Earth and Space Science Fellowship 80NSSC18K1268, the dissertation quarter fellowship from UC Santa Cruz, and 51 Pegasi b fellowship (Grant \#2023-4466) from the Heising-Simons Foundation. C.L. was supported by NASA's Juno project NNM06AA75C and sub-award to the University of Michigan with project No. Q99063JAR. X.Z. acknowledges support from the National Science Foundation grant AST2307463, NASA Exoplanet Research grant 80NSSC22K0236, and the NASA Interdisciplinary Consortia for Astrobiology Research (ICAR) grant 80NSSC21K0597. C.M. was supported by NASA’s Solar System Observations (SSO) award 80NSSC18K1001 to the University of California, Berkeley. H.G. acknowledges NASA's supercomputer Pleiades and supercomputer Lux at UC Santa Cruz, funded by NSF MRI grant AST 1828315.

\appendix

\section{Continuity Equations of Moisture, Clouds, and Precipitations}
\label{app:mass-flux}

The mass continuity equation of any species (e.g., vapor, clouds, precipitations) with density $\rho_{i}$ and chemical production and loss (e.g., phase transitions and cloud microphysics) can be generalized as
\begin{equation}
\label{eq:mass-continuity}
    \frac{\partial \rho_{i}}{\partial t} + \frac{1}{r \cos{\theta}}\frac{\partial (\rho_{i} u)}{\partial \phi} + \frac{1}{r \cos{\theta}} \frac{\partial (\rho_{i} v \cos{\theta})}{\partial \theta} + \frac{1}{r^2}\frac{\partial [\rho_{i} r^2 (w+V_{T})]}{\partial r} = \mathcal{P} - \mathcal{L},
\end{equation}
where $\theta$ is latitude; $\phi$ is longitude; $u,v,w$ are velocities in longitude, latitude, and radial directions, respectively; $V_{T}$ is the sedimentation velocity of a certain type of species, it is zero for vapor and cloud particles; $\mathcal{P}$ is the chemical production rate; $\mathcal{L}$ is the chemical loss rate. The averaged mass continuity equation is,
\begin{equation}
\label{eq:averaged-mass-continuity}
    \overline{\frac{\partial \rho_{i}}{\partial t}} + \overline{\frac{1}{r^2}\frac{\partial [\rho_{i} r^2 (w+V_{T})]}{\partial r}} = \overline{\mathcal{P}} - \overline{\mathcal{L}},
\end{equation}
where the terms on the LHS of Eq~\ref{eq:mass-continuity} about the flux divergence in the horizontal plane vanish. At the long-term-averaged statistical steady state, $\overline{\partial \rho_{i}/\partial t}\sim 0$. We further assume the radial geometric expansion in the weather layer is negligible as in Section~\ref{sec:self-regulation}, Eq~\ref{eq:averaged-mass-continuity} can be simplified as,
\begin{equation}
    \overline{\frac{\partial [\rho_{i} (w+V_{T})]}{\partial z}} = \overline{\mathcal{P}} - \overline{\mathcal{L}}.
\end{equation}

Then, we have a set of equations for vapor, clouds, and precipitations, 
\begin{equation}
\label{eq:averaged-vapor}
    \overline{\frac{\partial (\rho_{v} w)}{\partial z}} = \overline{\mathcal{P}_{\rm eva}} - \overline{\mathcal{L}_{\rm cond}},
\end{equation}

\begin{equation}
\label{eq:averaged-cloud}
    \overline{\frac{\partial (\rho_{c} w)}{\partial z}} = \overline{\mathcal{P}_{\rm cond}} - \overline{\mathcal{L}_{\rm c\rightarrow p}},
\end{equation}

\begin{equation}
\label{eq:averaged-precip}
    \overline{\frac{\partial [\rho_{p} (w+V_{T})]}{\partial z}} = \overline{\mathcal{P}_{\rm c\rightarrow p}} - \overline{\mathcal{L}_{\rm eva}}.
\end{equation}
Adding up Eq~\ref{eq:averaged-vapor} to \ref{eq:averaged-precip}, chemical reaction terms on the RHS vanish,

\begin{equation}
\label{eq:averaged-total}
    \frac{\partial [\overline{(\rho_{v}w)} + \overline{(\rho_{c}w)} + \overline{(\rho_{p}w)}] }{\partial z} = -\overline{\frac{\partial (\rho_{p}V_{T})}{\partial z}}.
\end{equation}
The mass flux of clouds and precipitation transported by the background velocity field is usually negligible compared with the mass flux of vapor since the density of clouds and precipitations is less than the mass flux of vapor. Then we can simplify Eq~\ref{eq:averaged-total} as,
\begin{equation}
    \overline{\rho_{v}w} \sim -\overline{\rho_{p}V_{T}}.
\end{equation}

\section{Table of Variables and Relations}

Table~\ref{tab:Vars} contains variables and useful relations used in the main text.

\begin{table}[t]
  \centering
  \caption{Thermodynamic variables and useful relations}
  \label{tab:Vars}
  \begin{tabular}{l|c|c|c}
    \toprule
    Variables/Relations & Symbols & Unit \\
    \midrule
    Gravitational acceleration & $g$ & $\rm m\;s^{-2}$ \\
    Pressure & $p$ & Pa \\
    Kinetic temperature & $T$ & K \\
    Effective temperature & $T_{\rm eff}$ & K \\
    Atmospheric density & $\rho$ & $\rm kg\;m^{-3}$ \\
    Molecular weight & $M$ & $\rm kg\; mol^{-1}$ \\
    Molecular weight of the hydrogen-helium mixture & $M_{d}$ & $\rm kg\; mol^{-1}$ \\
    Mean molecular weight & $\overline{M}$ & $\rm kg\; mol^{-1}$ \\
    Ratio of molecular weight of species $n$ to the dry component & $\epsilon_{n} = M_{n}/M_{d}$ & --- \\
    Universal gas constant & $R$ & $\rm J\;K^{-1}\;mol^{-1}$ \\
    Specific gas constant of species $n$ & $R_{n} = R/M_{n}$ & $\rm J\;K^{-1}\;kg^{-1}$ \\
    Pressure scale height & $H = \dfrac{RT}{\overline{M}g}$ & m \\
    Specific isochoric heat capacity & $c_{v}$ & $\rm J\;K^{-1}\;kg^{-1}$ \\
    Specific isobaric heat capacity & $c_{p}$ & $\rm J\;K^{-1}\;kg^{-1}$ \\
    Polytropic index & $\gamma$ & --- \\
    Specific latent heat & $L_{v}$ & $\rm J\;kg^{-1}$ \\
    Dimensionless specific latent heat of species $n$ & $\beta_{n} = \dfrac{L_{v,n}}{R_{n}T}$ & --- \\
    Dry adiabatic lapse rate & $\Gamma_{d}$ & $\rm K\;m^{-1}$ \\
    Moist adiabatic lapse rate & $\Gamma_{m}$ & $\rm K\;m^{-1}$ \\
    Saturation vapor pressure & $e_{s}$ & Pa \\
    Volume mixing ratio of vapor & $\eta$ & $\rm mol\; mol^{-1}$ \\
    Mass mixing ratio (mass fraction) & $q$ & $\rm kg\;kg^{-1}$ \\
    Brunt-V\"ais\"al\"a buoyancy frequency & $N$ & $\rm s^{-1}$\\
    Energy flux & $F$ & $\rm W\;m^{-2}$ \\
    Eddy-mean decomposition of quantity $A$ & $A = \overline{A}\;[\rm mean] + A'\;[\rm eddy]$ & --- \\
    Clausius-Clapeyron relation & $\dfrac{d\ln{e_{s,n}}}{dT} = \dfrac{L_{v,n}}{R_{n}T^2}$ & --- \\
    
    \bottomrule
  \end{tabular}
\end{table}

\section{The upper limit of the eddy diffusivity at the cloud level constrained by the heat flow}
\label{app-kzz}
In Section~\ref{sec:kzz}, we try to estimate the upper limit of the mixing efficiency at the cloud layer from the heat flow. Here, we derive the maximum $K_{zz}$ and show that it is determined by the local temperature, pressure, and relative humidity and independent of the atmospheric composition.

From Section~\ref{sec:kzz}, we know that,
\begin{equation}
\begin{split}
    K_{zz} & < -\frac{F_{\rm rad}^{\rm TOA}}{\rho L_{v}} \Big( \overline{\frac{d q_{v}}{d z}} \Big)^{-1}.
\end{split}        
\end{equation}
At the condensation level, the vertical gradient of the vapor mass mixing ratio is determined by the saturation vapor pressure and relative humidity. We assume that the relative humidity is 100\% to simplify the derivation. The gradient can be written as,
\begin{equation}
\begin{split}
    \overline{\frac{d q_{v}}{d z}} & = \frac{d}{dz}\Big[\frac{\eta_{s} M_{v}}{\eta_{s}M_{v}+(1-\eta_{s})M_{d}}\Big] \\
        & = \frac{M_{v}}{\eta_{s}M_{v}+(1-\eta_{s})M_{d}}\frac{d\eta_{s}}{d z} - \frac{M_{v}(M_{v}-M_{d})\eta_{s}}{[M_{v}\eta_{s}+(1-\eta_{s})M_{d}]^{2}}\frac{d\eta_{s}}{d z} \\
        & = \frac{M_{v}M_{d}}{\overline{M}^2}\frac{d\eta_{s}}{d z} \\
        & = \frac{M_{v}M_{d}}{\overline{M}^2}\frac{d}{d z}\Big( \frac{e_{s}}{p}\Big) \\
        & = \frac{M_{v}M_{d}}{\overline{M}^2}\Big( \frac{1}{p}\frac{d e_{s}}{d z}-\frac{e_{s}}{p^{2}}\frac{d p}{d z} \Big), \\
\end{split}
\end{equation}
From the Clausius-Clapeyron relation, we know that ${d\ln{e_{s}}}/{dT} = {M_{v}L}/{RT^2}$. Then, we can write the gradient as,

\begin{equation}
\begin{split}
    \overline{\frac{d q_{v}}{d z}} & = \frac{M_{v}M_{d}}{\overline{M}^2}\Big( \frac{e_{s}M_{v}L_{v}}{pRT^{2}}\frac{dT}{dz} + \frac{e_{s}\rho g}{p^{2}} \Big) \\
    & = \frac{M_{v}M_{d}}{\overline{M}^2}\Big( \frac{e_{s}\beta}{pT}\frac{dT}{dz}+\frac{\eta_{s}}{H} \Big).
\end{split}
\end{equation}
In ice giant and most giant planet atmospheres, $\eta_{s} = e_{s}/p \ll 1$, we can also use the dry adiabatic lapse rate to approximate the temperature gradient. Then, we can simplify the gradient of mass mixing ratio as,

\begin{equation}
\begin{split}
    \overline{\frac{d q_{v}}{d z}} & \sim \epsilon\Big( -\frac{\eta_{s}\beta g}{c_{p}T} + \frac{\eta_{s}}{H} \Big) \\
    & \sim \frac{\epsilon\eta_{s}}{H}\Big[ -\frac{(\gamma-1)\beta}{\gamma}+1 \Big].
\end{split}
\end{equation}

If we do not assume the atmosphere is saturated and the temperature profile does not follow dry adiabatic, $dT/dz = c\Gamma_{d}$ (i.e., if the weather layer is superadiabatic, then $c > 1$ and subadiabatic weather layer is $c < 1$),

\begin{equation}
    \overline{\frac{d q_{v}}{d z}} \sim \epsilon\eta_{s} \frac{\rm RH}{H}\Big[ -\frac{(\gamma-1)\beta c}{\gamma}+1 \Big] + \epsilon\eta_{s}\frac{d}{dz}{\rm RH} ,
\end{equation}

\bibliography{main}


\end{CJK}
\end{document}